\documentclass[
a4paper,reprint,
superscriptaddress,
groupedaddress,
nofootinbib,
bibnotes,
amsmath,amssymb,
aps,
floatfix]{revtex4-1}

\usepackage{graphicx}% Include figure files
\usepackage{dcolumn}% Align table columns on decimal point
\usepackage{bm}% bold math
\usepackage[mathlines]{lineno}% Enable numbering of text and display math
\usepackage{csquotes}
\usepackage{gensymb}
\usepackage{rotating}
\newcommand\ddfrac[2]{\frac{\displaystyle #1}{\displaystyle #2}}
\usepackage[dvipsnames]{xcolor}
\usepackage{multirow}
\usepackage{siunitx}
\usepackage[breaklinks=true]{hyperref}%hidelinks,%Add hypertext capabilities
\hypersetup{
    colorlinks,
    linkcolor={red!50!black},
    citecolor={blue!50!black},
    urlcolor={blue!80!black}
}

\begin{document}

\title{Muon deficit in air shower simulations estimated from AGASA \\ muon measurements}% Force line breaks with \\
\author{F.~Gesualdi}
\email{flavia.gesualdi@iteda.cnea.gov.ar}
\affiliation{Instituto de Tecnolog\'ias en Detecci\'on y Astropart\'iculas (CNEA, CONICET, UNSAM),
Centro At\'omico Constituyentes, San Mart\'in, Buenos Aires, Argentina\\}
\affiliation{Karlsruhe Institute of Technology, Institut f\"ur Kernphysik (IKP), Germany\\}

\author{A.~D.~Supanitsky}
\affiliation{Instituto de Tecnolog\'ias en Detecci\'on y Astropart\'iculas (CNEA, CONICET, UNSAM),
Centro At\'omico Constituyentes, San Mart\'in, Buenos Aires, Argentina\\}

\author{A.~Etchegoyen}
\affiliation{Instituto de Tecnolog\'ias en Detecci\'on y Astropart\'iculas (CNEA, CONICET, UNSAM),
Centro At\'omico Constituyentes, San Mart\'in, Buenos Aires, Argentina\\}

\date{\today}% It is always \today, today,
             %  but any date may be explicitly specified

\begin{abstract}

In this work, direct measurements of the muon density at $1000\,\textrm{m}$ from the shower axis obtained by the Akeno Giant Air Shower Array (AGASA) are analysed. The selected events have zenith angles $\theta \leq 36\degree$ and reconstructed energies in the range $18.83\,\leq\,\log_{10}(E_{R}/\textrm{eV})\,\leq\,19.46$. These are compared to the predictions corresponding to proton, iron, and mixed composition scenarios obtained by using the high-energy hadronic interaction models EPOS-LHC, QGSJetII-04, and Sibyll2.3c. The mass fractions of the mixed composition scenarios are taken from the fits to the depth of the shower maximum distributions performed by the Pierre Auger Collaboration. The cross-calibrated energy scale from the \textit{Spectrum Working Group} [D. Ivanov, for the Pierre Auger Collaboration and the Telescope Array Collaboration, PoS(ICRC2017) 498 (2017)] is used to combine results from different experiments. The analysis shows that the AGASA data are compatible with a heavier composition with respect to the one predicted by the mixed composition scenarios. Interpreting this as a muon deficit in air shower simulations, the incompatibility is quantified. The muon density obtained from AGASA data is greater than that of the mixed composition scenarios by a factor of $1.49\pm0.11\,\textrm{(stat)}\pm0.18\,\textrm{(syst)}$, $1.54\pm0.12\,\textrm{(stat)}\pm0.18\,\textrm{(syst)}$, and $1.66\pm0.13\,\textrm{(stat)}\pm0.20\,\textrm{(syst)}$ for EPOS-LHC, Sibyll2.3c, and QGSJetII-04, respectively.

\end{abstract}

\pacs{}% PACS, the Physics and Astronomy
                             % Classification Scheme.
%\keywords{Suggested keywords}%Use showkeys class option if keyword
                              %display desired
\maketitle

\allowdisplaybreaks %it is good for removing empty spaces

\section{Introduction}

Although in recent years a significant progress in the study of ultra-high-energy cosmic rays (UHECRs) has been achieved, essential 
aspects remain unresolved. Among the open questions are \enquote{where do they come from?}, \enquote{how do they accelerate to the 
highest energies?}, and \enquote{what is its nature?}. To answer those questions, the UHECRs are studied through the measurement of the energy spectrum, the distribution of their arrival directions, and the primary mass composition as a function 
of the energy. 

Because the flux drops steeply, cosmic rays with energies above $10^{15}\,\textrm{eV}$ can only be studied through large ground based
observatories, which provide enough exposure for the detection of extensive air showers (EASs). The latter consist of billions of secondary
particles resulting from the interaction of the primary cosmic ray with the atmosphere. Each EAS can be divided into three components: the
hadronic, the muonic, and the electromagnetic. The hadronic component, mostly consisting of neutral and charged pions, protons, antiprotons,
and neutrons, feeds the muonic and electromagnetic components. The latter is composed by electrons, positrons and photons, and is the 
dominant component as it carries most of the energy of the shower. The muonic component, comprised of muons and antimuons, originates 
mainly from the decay of hadrons (only a very small fraction is produced from the electromagnetic component), and therefore serves as a 
tracer of the hadronic component because most of these particles reach the detectors before decaying. 

The energies to which cosmic rays can reach are inaccessible at the Large Hadron Collider (LHC). This opens the door to testing high-energy hadronic interaction models at ultra-high energies. Recently, the most widely used 
models have been updated to LHC data. They are QGSJetII-04 \cite{QGSJet}, EPOS-LHC \cite{Epos}, and Sibyll2.3c \cite{Sibyll}. These 
are referred to as post-LHC models due to their tuning to LHC data.

UHECRs are known to be predominantly nuclei ranging from proton (light) to iron (heavy) \cite{Book}. These 
charged nuclei are deflected by magnetic fields as they propagate from their sources to the Earth's atmosphere. Since light nuclei 
are less deflected than heavy ones, the primary nature is of crucial importance for the identification of the sources, which
could be possible considering the light component at the highest energies \cite{AugerPrime:16}. Furthermore, composition information 
is also important to understand the transition between galactic and extra-galactic cosmic rays \cite{Book}.

The EAS observables most sensitive to the nature of the primary are the depth of the shower maximum $X_{\textrm{max}}$ and the number of muons produced in the shower, or equivalently, the muon density $\rho_\mu$ at a given distance to the shower axis.

It is known that the mean of $X_{\textrm{max}}$, denoted as $\langle X_{\textrm{max}} \rangle$, is smaller for heavier primaries because their first interaction occurs higher in the 
atmosphere, and also because the generated EASs develop faster compared to the ones generated by lighter primaries \cite{Book}. Due to its
primary mass sensitivity, $X_{\textrm{max}}$ is commonly used for composition analyses by fitting the energy-binned measured
$X_{\textrm{max}}$ distributions with a linear combination of, for example, four single nuclei simulated $X_{\textrm{max}}$ 
distributions \cite{Bellido18}. 

EAS simulations that make use of post-LHC models reproduce to a good extent the behaviour of the $X_{\textrm{max}}$ parameter. The 
predicted $\langle X_{\textrm{max}}\rangle$ and mean-logarithmic-mass $\langle \ln A \rangle$ differ in $\sim \pm 0.8$ in 
$\langle \ln A \rangle$ between models and the difference is fairly constant as a function of the primary energy. Furthermore, 
the theoretical uncertainties of $X_{\textrm{max}}$ are relatively small compared to those of $\rho_\mu$ \cite{Pierog17,Prado18}. 
For these reasons, it is customary to test other EAS observables by comparing their composition interpretation to the one obtained 
from $X_{\textrm{max}}$. Inconsistent interpretations would imply that the models do not reproduce properly all EAS observables.

A muon deficit in interaction models has been reported by numerous collaborations. A combined analysis of eight experiments (EAS-MSU, 
IceCube Neutrino Observatory, KASCADE-Grande, NEVOD-DECOR, Pierre Auger Observatory, SUGAR, Telescope Array and Yakutsk) shows that
simulations and muon measurements are consistent up to $10^{16}\,\textrm{eV}$ \cite{Dembinski19,Cazon19likeDembinski19}. However, at higher 
energies the deficit is found to increase with the energy. The discrepancy is smaller for the updated models \cite{Dembinski19}. Furthermore, 
the muon deficit is greater for larger values of the zenith angle \cite{MuonDeficit16} and at larger distances to the shower axis
\cite{TAmuons}. 

Three different experiments studied the muon deficit in an energy range which overlaps with the one of this work: First, the Pierre Auger Observatory reported a muon deficit of $30\,\%$ to $80\,\%$ in the mixed composition scenarios  \cite{MuonDeficit15,MuonDeficit16}. Second, the Telescope Array Collaboration observed a deficit of $\sim\!67\,\%$ against proton-induced QGSJetII-04 simulations, the latter being in agreement with the composition derived from their $X_{\textrm{max}}$ measurements \cite{TAmuons}. Finally, Yakutsk data suggest lower muon densities which are compatible with no muon deficit \cite{Dembinski19,Cazon19likeDembinski19}. 
At lower energies, with AMIGA (the muon detectors of a low-energy extension of the Pierre Auger Observatory) the deficit is found to be between $38\,\%$ and $53\,\%$ ($10^{17.5}\,\textrm{eV}\lesssim E\lesssim10^{18.0}\,\textrm{eV}$) \cite{Muller18}. 
In addition, HiRes/MIA 
($10^{17}\, \textrm{eV}\lesssim E\lesssim10^{18}\, \textrm{eV}$), and NEVOD-DECOR
($10^{15}\,\textrm{eV}\lesssim E\lesssim10^{18}\,\textrm{eV}$)
%, and Yakutsk ($10^{17}\,\textrm{eV}\lesssim E\lesssim10^{19}\,\textrm{eV}$) 
experiments reported a muon deficit in the specified energy ranges. 
In contrast, the EAS-MSU ($10^{17}\,\textrm{eV}\lesssim E\lesssim10^{18}\,\textrm{eV}$), the IceCube Neutrino Observatory 
($10^{15}\,\textrm{eV} \lesssim E\lesssim10^{17}\,\textrm{eV}$), and KASCADE-Grande ($E\sim10^{17}\,\textrm{eV}$) reported no muon 
deficit in the energy range on which they operate (see Refs.~\cite{Dembinski19,Cazon19likeDembinski19} and references therein). It should be noted that the uncertainties in the energy scales of the experiments are non-negligible, and translate almost directly into uncertainties in the data to Monte Carlo ratio of muon density or muon number \cite{Dembinski19,Cazon19likeDembinski19}.

It remains unclear whether the muon deficit is originated by a new phenomenon at high energies or by a partial mismodelling of hadronic
collisions at high or low energies \cite{MuonDeficit16}. Understanding the muon deficit would allow the models to reproduce more faithfully 
the behaviour of EASs, reducing the systematic uncertainties of mass composition analyses.

In this work, muon density measurements from the Akeno Giant Air Shower Array (AGASA) are used to study the muon deficit in air shower
simulations. The AGASA experiment consisted of an array of 111 scintillation counters spread across $\sim\! 100\,\textrm{km}^{2}$, as well as 
27 muon detectors. The latter were formed by proportional counters shielded with $30\,\textrm{cm}$ of iron or $1\,\textrm{m}$ of 
concrete (the vertical muon energy threshold was $0.5\,\textrm{GeV}$). The experiment was able to measure events with energies above $3 \times 10^{16}\,\textrm{eV}$ and with zenith angles $\theta \leq 45\degree$ \cite{AGASA}. The detectors were decommissioned in 2004.

The data set under analysis is particularly relevant because the hybrid design of AGASA allows for the measurement of primary 
energy and, simultaneously, the direct detection of muons at energies above $10^{19}\,\textrm{eV}$. 
The determination of the muon deficit from AGASA data is complementary to other measurements as it explores another region of the parameter 
phase-space, and contributes to reducing the overall uncertainties.

The article is organised as follows. In Sec.~\ref{secAnalysis} a description of the analysis is presented, which includes:
the development of a method to take into account the effects of the energy reconstruction in simulations, the transformations of the energy scales of the different experiments relevant to this work to the reference energy scale proposed by the \textit{Spectrum Working Group} \cite{Ivanov17}, and the calculation of the average muon density divided by the energy from data, simulations, and for the mixed composition scenarios which combine both of them. In Sec.~\ref{secResults} the results are presented and in Sec.~\ref{secConc} the main conclusions are summarised.

\section{Analysis}\label{secAnalysis}

\subsection{Effect of the reconstructed energy uncertainty on the muon density}\label{sec1}

Whereas the simulated muon density at $1000\,\textrm{m}$ from the shower axis is a function of the true or 
input energy $E$, the measured muon density is a function of the reconstructed energy $E_R$. Therefore, a straightforward 
comparison is not appropriate, even if $E_R$ is an unbiased estimator of $E$ \cite{Dembinski18}. 

The simulated average muon density divided by the reconstructed energy, calculated in the 
$i$-th reconstructed energy bin, takes the following form,

\begin{widetext}
\begin{equation}
\left\langle\ddfrac{\rho_{\mu}}{E_R}\right\rangle(E_{Ri}) = \ddfrac{\int_{E_{Ri}^{-}}^{E_{Ri}^{+}} \int_{0}^{\infty}
\langle \widetilde{\rho}_{\mu}\rangle(E) \, E_R^{-1} \, J(E)\, G(E_{R}|E)dE\, dE_{R}}{\int_{E_{Ri}^{-}}^{E_{Ri}^{+}} 
\int_{0}^{\infty} J(E)\, G(E_{R}|E)dE\, dE_{R}},
\label{EqRhoAv}
\end{equation}
\end{widetext}
where $E_{Ri}$ is the center of the reconstructed energy bin, $E_{Ri}^{-}$ and $E_{Ri}^{+}$ are the lower and upper limits of that bin. 
Here,
\begin{itemize}

\item $\langle \widetilde{\rho}_{\mu}\rangle (E)$ is the average muon density as a function of the true or input energy of the simulation, which is obtained from fits to shower simulations that are performed by using CORSIKA version 7.6400 \cite{CORSIKA} (the tilde is to emphasize that this quantity is not directly comparable to the average muon density computed from data);

\item $J(E)$ is the cosmic ray flux, which is obtained by fitting the Telescope Array measurements with an appropriate
function \cite{TAFlux};

\item $G (E_{R}|E)$ is the conditional probability distribution of $E_R$ conditioned to $E$, which is reported to be a 
log-normal distribution \cite{Takeda2003} with a standard deviation that decreases with energy \cite{Yoshida95}. 

\end{itemize}
The details about the determination of these functions are given in Appendix \ref{secAppFunctions}.

The rationale behind Eq.~(\ref{EqRhoAv}) is the following. The energy of a real or simulated air shower with true energy $E$ is 
estimated by means of the reconstruction procedure producing a value, $E_R$, according to $G(E_R|E)$. Furthermore, the distribution 
of the true energy $E$ is given by the cosmic ray flux $J(E)$ (normalized within a certain energy range). The product $J(E) G(E_R|E)$ 
represents the joint probability distribution of $E$ and $E_R$. While $G(E_R|E)$ can be thought of, roughly, as a Gaussian-like 
distribution, $J(E)$ is highly asymmetric as it drops steeply with energy. Therefore, the product $J(E) G(E_R|E)$ is asymmetric too, 
being higher for lower energies. In other words, an event with reconstructed energy $E_R$ can come, most likely, from an event that 
has a true energy $E$ smaller than $E_R$. The mean value $\langle\rho_{\mu}/E_R \rangle (E_R)$ can be
calculated via the integration of the contributions of $\langle\widetilde{\rho}_{\mu} \rangle (E)  /E_R$ weighted by the product 
$J(E) G(E_R|E)$ (again, normalized within a certain energy range). Finally, the integration in a reconstructed energy bin is 
introduced, taking it into account in the normalization as well.

$\langle\widetilde{\rho}_{\mu} \rangle(E)$ is essentially a power law in energy, i.e.~$\propto E^{\beta}$, with $\beta \sim 0.9$.
Therefore, $\langle\widetilde{\rho}_{\mu} \rangle(E)$ is smaller for lower energies. As explained before, lower energies weigh more 
in the integration. It follows that, evaluated at a specific numerical value $E^*$, $\langle \rho_{\mu} /E_R \rangle (E_R=E^*) < \langle
\widetilde{\rho}_{\mu}/E \rangle (E=E^*)$. The difference increases for broader conditioned distributions $G(E_{R}|E)$ and in regions 
where the flux $J(E)$ is steeper. An additional (though smaller) effect is introduced from the binning in reconstructed energy: if the 
bin is centered at $E_{Ri}$, then $\langle\rho_{\mu}/E_R \rangle (E_{Ri}=E^*) < \langle\rho_{\mu}/E_R \rangle (E_R=E^*)$.

\begin{figure}[!ht]
\includegraphics[width=\linewidth]{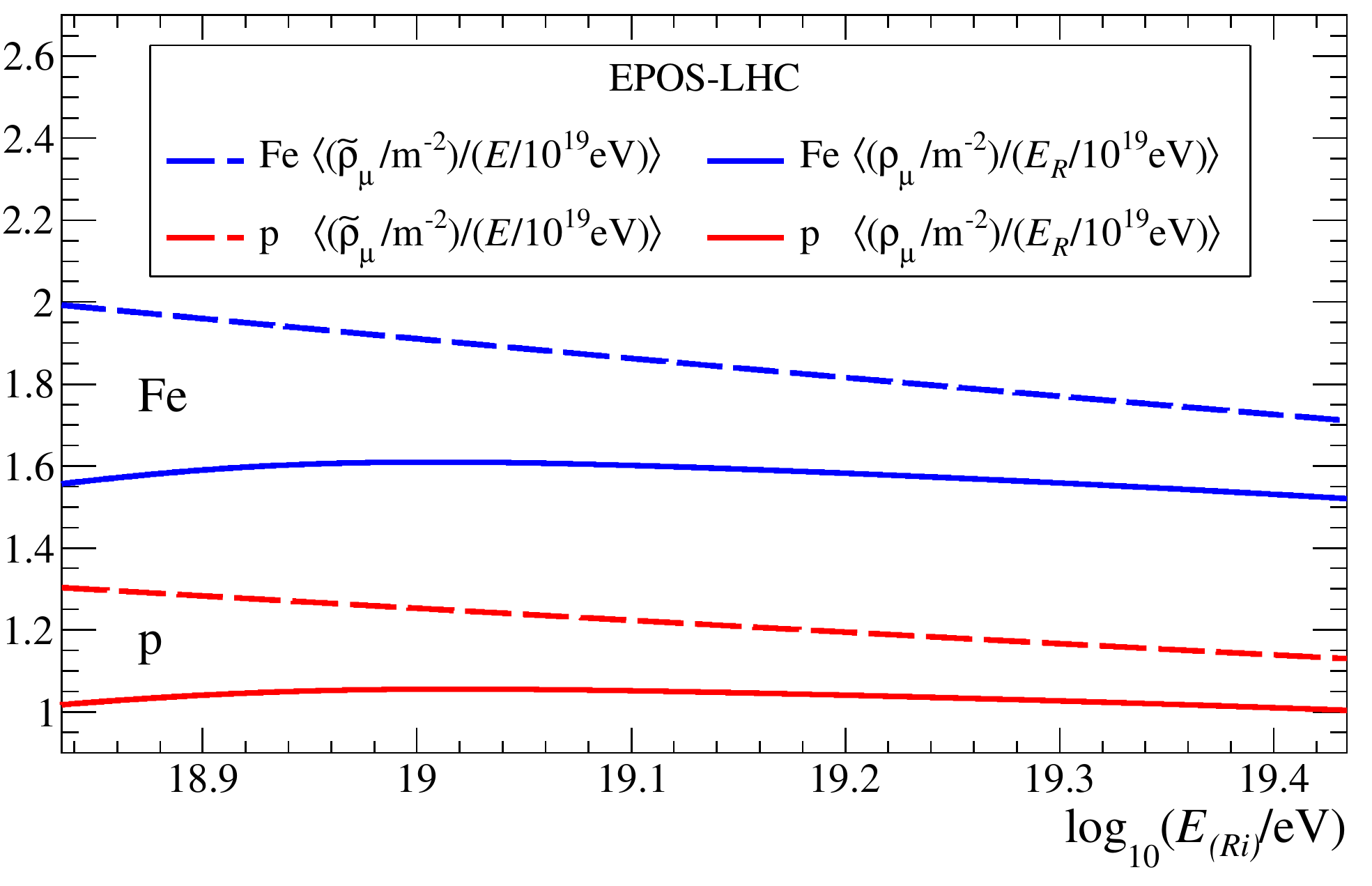}
\caption{Average muon density at $1000\,\textrm{m}$ divided by the energy (reconstructed energy), as a function of the logarithm of the input energy of the 
simulations (the logarithm of the reconstructed energy in the center of the $i$-th bin) in dashed lines (solid lines). The 
bin width considered is $\Delta\log_{10}(E/\textrm{eV})=0.2$. The model used is EPOS-LHC and the primaries are proton (red) and iron (blue). 
\label{fconv}}
\end{figure}

Figure \ref{fconv} shows a comparison between $\langle\widetilde{\rho}_{\mu}/E\rangle(E)$ and $\langle\rho_{\mu}/E_R\rangle(E_{Ri})$. From 
the figure it can be seen that $\langle\rho_{\mu}/E_R\rangle(E_{Ri})$ can be $11\,\%$ to $22\,\%$ smaller than 
$\langle \widetilde{\rho}_{\mu}/E\rangle(E)$ in the analysed energy range, i.e.~from $10^{18.83}\,\textrm{eV}$ to 
$10^{19.46}\,\textrm{eV}$. At low energies, this difference is explained by the large uncertainty 
in the reconstructed energy ($\sim\! 28\,\%$ at $10^{18.83}\,\textrm{eV}$). At high energies, the dominant effect 
is the flux suppression. The effect of the binning in reconstructed energy with a bin width of $\Delta\log_{10}(E_R/\textrm{eV})=0.2$ 
is small in comparison to the one introduced by the energy uncertainty in combination with the flux shape.

It is relevant to add that, in practice, none of the functions in Eq.~(\ref{EqRhoAv}) are defined from $0$ to $\infty$ in $E$. 
The integration range is limited to the smallest definition range of all functions, which is that of 
$\langle \widetilde{\rho}_{\mu} \rangle(E)$, i.e.~the one corresponding to the simulations ($18.0\,<\,\log_{10}(E/\textrm{eV})\,<\,19.8$). 
The effect on $\langle\rho_{\mu}/E_R\rangle(E_{Ri})$ of taking a small integration range instead of the infinite one is estimated to be of 
$\sim\!0.1\,\%$ in the analysed energy range, which is negligible compared to the other uncertainties. The integrals are performed 
numerically by using ROOT \cite{ROOT}.

\subsection{Transformation to the reference energy scale} \label{sec2}

In this work, data from three different experiments are used: the AGASA muon density as a function of the energy, the 
Telescope Array cosmic ray energy spectrum, and the Auger mass composition fractions as a function of the primary energy 
(obtained by fitting the $X_{\textrm{max}}$ experimental distributions). 

These three experiments have different energy scales $E^{\textrm{data}}$. Therefore, these scales are shifted by a factor 
$f_{\textrm{E}} = E^{\textrm{ref}}/E^{\textrm{data}}$ to bring them to the cross-calibrated energy scale $E^{\textrm{ref}}$ 
found by the \textit{Spectrum Working Group} \cite{Ivanov17}. The $f_{\textrm{E}}$ factors are found by matching flux 
measurements, based on the assumption that the cosmic ray flux is isotropic and therefore should be the same for all 
experiments.

The factors $f_{\textrm{E}}$ and the relative systematic uncertainties of the energy in the original 
($\varepsilon_{\textrm{SD}} = \sigma^{\textrm{syst}}(E^{\textrm{data}})/E^{\textrm{data}}$) and reference energy scale 
($\varepsilon_{\textrm{SR}} = \sigma^{\textrm{syst}}(E^{\textrm{ref}})/E^{\textrm{ref}}$) are reported in Table \ref{Table1}. 
The $f_{\textrm{E}}$ values for Auger and Telescope Array are taken from Ref.~\cite{Ivanov17}. The value of $f_{\textrm{E}}$ for 
AGASA is given by,
\begin{equation}
f_{\textrm{E}}=\frac{E^{\textrm{ref}}}{E^{\textrm{TA}}} \times \frac{E^{\textrm{TA}}}{E^{\textrm{AGASA}}} = 0.948 \times 0.72 = 0.68, 
\end{equation}
where $E^{\textrm{TA}}/E^{\textrm{AGASA}}$ is taken from Ref.~\cite{doctesisIvanov}. The relative systematic uncertainties, 
$\varepsilon_{\textrm{SD}}$, of Auger, Telescope Array, and AGASA are taken from Refs.~\cite{EScaleAuger},~\cite{EScaleTA},
and \cite{Takeda2003}, respectively. 

Given an energy value $E_0^{\textrm{data}}$ measured in the energy scale of a certain experiment, it is imposed that the energy values in the interval $[(1-\varepsilon_{\textrm{SD}}^-),(1+\varepsilon_{\textrm{SD}}^+)] \times E_0^{\textrm{data}}$ measured in the original scale are also the values enclosed by the corresponding interval in the reference scale, i.e.~$[(1-\varepsilon_{\textrm{SR}}^-),(1+\varepsilon_{\textrm{SR}}^+)]\times E_0^{\textrm{ref}}$, where $E_0^{\textrm{ref}} = f_E \, E_0^{\textrm{data}}$. This leads to the following expressions for the upper and lower boundaries of the relative systematic uncertainties corresponding to the reference energy scale,
\begin{eqnarray}
\label{eqSRplus}
1 + \varepsilon_{\textrm{SR}}^+ = (1 + \varepsilon_{\textrm{SD}}^+) \times \frac{1}{f_{\textrm{E}}},\\
\label{eqSRminus}
1 - \varepsilon_{\textrm{SR}}^- = (1 - \varepsilon_{\textrm{SD}}^-) \times \frac{1}{f_{\textrm{E}}}.
\end{eqnarray}
From Eqs.~(\ref{eqSRplus}) and (\ref{eqSRminus}) it is easy to understand how a symmetric systematic uncertainty in the original energy scale becomes an asymmetric systematic uncertainty in the reference energy scale.  
\begin{table}[h]
\begin{ruledtabular}
\begin{tabular}{cccc}
%\hline
Observatory &  $f_{\textrm{E}}$ & $\varepsilon_{\textrm{SD}}$ & $\varepsilon_{\textrm{SR}}$  \\[0.2cm]
 \hline
 \\[-0.2cm]
 Pierre Auger     & 1.052 & $\pm14\,\%$ & $^{+8.4}_{-18}\,\%$ \\ [0.2cm]
 Reference        & 1 & $\pm10\,\%$ & $\pm10\,\%$   \\ [0.2cm]
 Telescope Array  & 0.948 & $\pm21\,\%$ & $^{+28}_{-17}\,\%$ \\ [0.2cm] 
 AGASA            & 0.68  & $\pm18\,\%$ & $^{+72}_{+20}\,\%$ \\ [0.2cm] 
\end{tabular}
\end{ruledtabular}
\caption{Energy scale correction factors, obtained from the cross-calibration of the flux measurements, and relative systematic 
uncertainties of the energy in the original and reference energy scales.}
\label{Table1}
\end{table}

The relative systematic uncertainty of the reference energy scale is reported to be of at least $10\,\%$ \cite{Dembinski19,Cazon19likeDembinski19}, 
which is the value adopted in this work. In any case, in Section \ref{secResults} it is discussed how the results are affected 
by taking the largest systematic uncertainties given by Telescope Array and Auger ($^{+28}_{-18}\,\%$, see Table \ref{Table1}). 
The systematic uncertainties on the energy scale of AGASA are not taken into account since they are incompatible with the 
reference energy scale.

\subsection{Calculation of the muon density} \label{sec3}

\textbf{Data:} Muon density in the analysed AGASA measurements is determined as the so-called \enquote{on-off density} \cite{Shinozaki2004}. This is computed by using the number of segments that were hit $n$ out of the total available ones $m$ within one detector of area $A$. Assuming a Poissonian distribution, the muon density is $\rho_{\mu} = -m \ln(1-n/m)/A$. This is a good estimator provided that showers are nearly-vertical and that muon densities are $\lesssim 10\,\textrm{m}^{-2}$ (such that $n\ll m$) \cite{AGASA,Shinozaki2004}. Then, the muon density at $1000\,\textrm{m}$ from the shower axis is determined from the fit of the measurements to a muon lateral distribution function \cite{AGASA}; its uncertainty is reported to be $\sim\! 40\,\%$ above $10^{19}\,\textrm{eV}$ (see Ref.~\cite{Shinozaki2004} and references therein). The muon density values of the analysed events are extracted from Fig.~7 of Ref.~\cite{Shinozaki2004}\footnote{Previous versions of this data set can be found in Refs.~\cite{Shinozaki2001,Shinozaki2002}.}; they are shown in Fig.~\ref{fRawData}, and are also listed in Appendix \ref{secAppRawData}. The data set consists of events restricted to zenith angles $\theta \leq 36\degree$, with a vertical muon energy threshold of $0.5\,\textrm{GeV}$ \cite{Shinozaki2004}. The events with no muon detection, below the dashed line in Fig.~\ref{fRawData}, are included in the analysis. The energy cut at $\log_{10}(E_{R}/\textrm{eV})=19.46$ is set due to the sharp drop in statistics beyond that energy.

\begin{figure}[!ht]
\includegraphics[width=\linewidth]{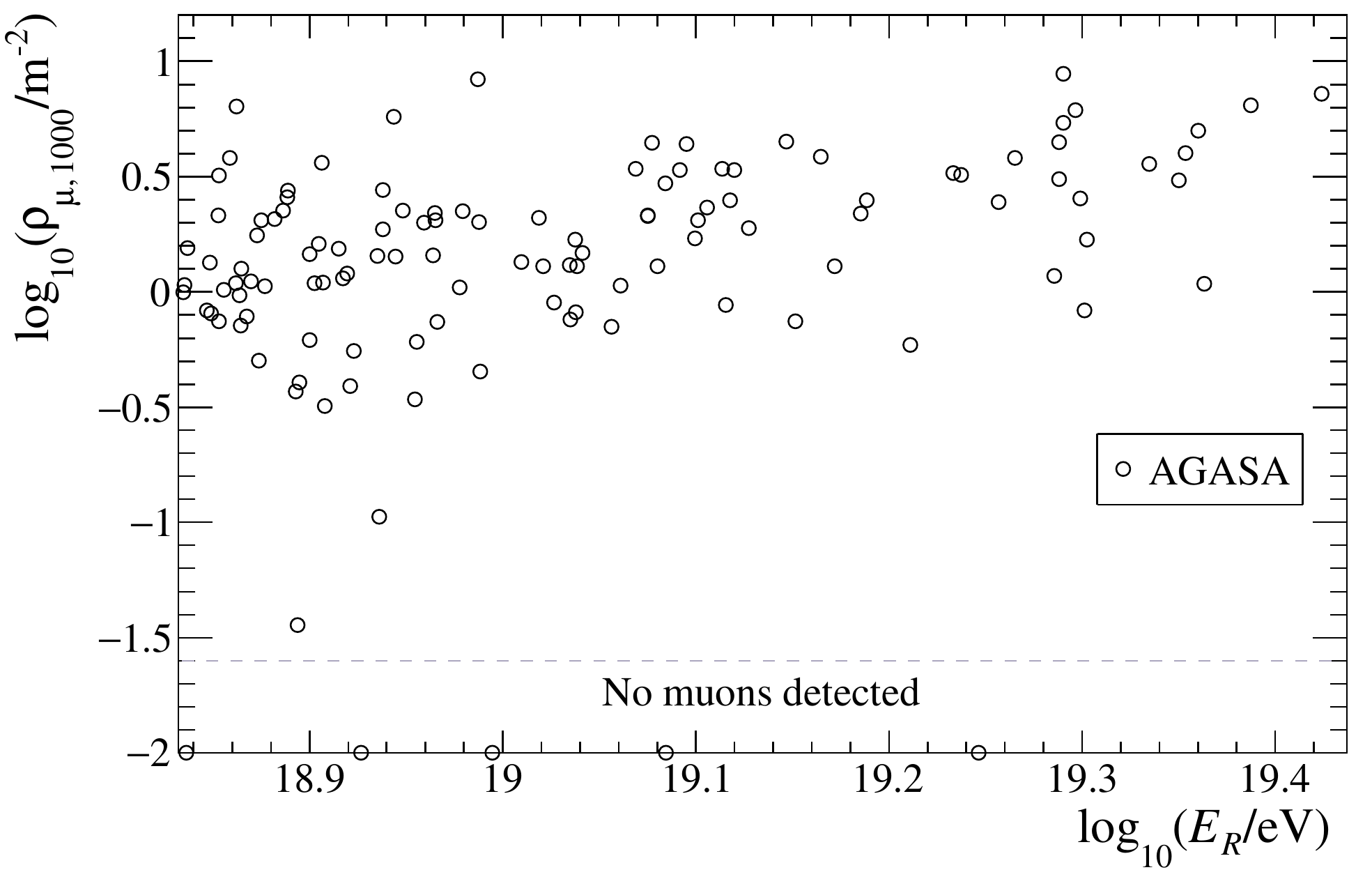}
\caption{Logarithm of the muon density as a function of the logarithm of the reconstructed energy (in the reference scale). For the events below the dashed line no muons were measured in any muon detector. The data points are extracted from Fig.~7 of Ref.~\cite{Shinozaki2004}.\label{fRawData}}
\end{figure}

The energy of an event in AGASA is estimated through a function which depends almost linearly on $S_{0}(600)$. This is the density of charged
particles at $600\,\textrm{m}$ from the shower axis obtained from the fit of the experimental lateral distribution function, normalized 
at $0 \degree$ zenith angle \cite{Takeda2003}. The explicit conversion formula is reported in Appendix \ref{secAppFunctions}.

\textbf{Simulations:} Proton, helium, nitrogen and iron initiated air showers are simulated for the models QGSJetII-04, EPOS-LHC 
and Sibyll2.3c, and low-energy hadronic interaction model Fluka version 2011.2x \cite{Fluka1,Fluka2}. For each model and primary type, $\sim\!20$ 
showers ($\sim\!30$ for proton primaries) per input energy are simulated, in the energy range $18.0\,\leq\,\log_{10}(E/\textrm{eV})\,\leq\,19.8$ and in steps of $\Delta\log_{10}(E/\textrm{eV})\,=\,0.2$. It is worth mentioning that a larger number of proton-initiated showers (with respect to iron-initiated 
showers) are simulated because shower-to-shower fluctuations are larger for lighter primaries. Furthermore, additional showers for proton and iron primaries of models QGSJetII-04 and EPOS-LHC are simulated in the energy range $19.8\,\leq\,\log_{10}(E/\textrm{eV})\,\leq\,20.8$ to validate the performance of the integral in Eq.~(\ref{EqRhoAv}) in a finite energy range. 
Some relevant parameters of the simulations are given in Appendix \ref{secAppParams}.

From every simulated air shower, the muon density is estimated by counting the muons in a $10\,\textrm{m}$ wide ring of $1000\,\textrm{m}$
radius, measured in the shower plane. For a given input energy value, the muon density of the $\sim\! 20$ (or $\sim\! 30$) showers 
is averaged and the standard deviation of the mean is taken as its statistical uncertainty. The average muon density at $1000\,\textrm{m}$
from the shower axis as a function of the input energy is obtained by fitting the simulated data with a power law in $E$,
\begin{equation}
\langle\widetilde{\rho}_{\mu,\,1000}\rangle(E) = \rho_{\mu (19)} \left(\frac{E}{10^{19}\,\textrm{eV}}\right)^{\beta},
\label{eqLinFit}
\end{equation}
where $\rho_{\mu (19)}$, the muon density at $10^{19}\,\textrm{eV}$, and $\beta$ are free fit parameters. This is done for all 
models and primary types. The results of the fits are given in Appendix \ref{secAppFunctions}. 

As mentioned before, $\langle\rho_{\mu,\,1000}/E_R\rangle(E_{Ri})$ is obtained from $\langle\widetilde{\rho}_{\mu,\,1000}\rangle(E)$ via the numerical evaluation of Eq.~(\ref{EqRhoAv}).

\textbf{Mixed composition scenarios:} Muon densities for the mixed composition scenarios 
$\langle\widetilde{\rho}_{\mu,\,1000}^{\ \textrm{mix}}\rangle(E)$ are derived by using the mass fractions obtained by the Pierre Auger Collaboration 
from the fits to the $X_{\textrm{max}}$ experimental distributions (see Ref.~\cite{Bellido18} for details). Therefore, for each 
model the muon density is given by,
\begin{equation}
\langle\widetilde{\rho}_{\mu,\,1000}^{\ \textrm{mix}}\rangle(E) = \sum_{A} f_A(E) \, \langle\widetilde{\rho}_{\mu,\,1000}^{A}\rangle(E),
\label{eqmix}
\end{equation}
where $A=\{\textrm{p, He, N, Fe} \}$ and $f_A(E)$ is the mass fraction as a function of primary energy, obtained by transforming
the Auger energy to the one corresponding to the reference energy scale. 

$\langle\rho_{\mu,\,1000}^{\textrm{mix}}/E_R\rangle(E_{Ri})$ is calculated from $\langle\widetilde{\rho}_{\mu,\,1000}^{\ \textrm{mix}}\rangle(E)$ through 
Eq.~(\ref{EqRhoAv}). The mass fractions obtained by Auger are given for discrete values of primary energy. Therefore, the 
integration in the variable $E$ of Eq.~(\ref{EqRhoAv}) is performed considering a linear interpolation of the mass fractions
values. It is worth mentioning that $\langle\rho_{\mu,\,1000}^{\textrm{mix}}/E_R\rangle(E_{Ri})$ does not result in a linear combination of 
$\langle\rho_{\mu,\,1000}^{A}/E_R\rangle(E_{Ri})$ since the mass fractions $f_A(E)$ depend on the energy $E$, which is an integration variable.

The statistical and systematic uncertainties of $\langle\rho_{\mu,\,1000}^{\textrm{mix}}/E_R\rangle(E_{Ri})$ are assessed as follows: for 
a certain interaction model, for each discrete energy value, the combination of mass fractions within the boundaries of its uncertainties that maximize 
and minimize $\langle\widetilde{\rho}_{\mu,\,1000}^{\ \textrm{mix}}\rangle(E)$ are selected (this is an over-estimation, but this method is the best approach given that the covariance matrices of the mass fraction fits are not available). 
In this way, $\langle\widetilde{\rho}_{\mu,\,1000}^{\ \textrm{mix}}\rangle(E) \pm \sigma[\langle\widetilde{\rho}_{\mu,\,1000}^{\ \textrm{mix}}\rangle](E)$ is calculated for each discrete energy value. Subsequently, the values of $\langle\widetilde{\rho}_{\mu,\,1000}^{\ \textrm{mix}}\rangle(E) + \sigma[\langle\widetilde{\rho}_{\mu,\,1000}^{\ \textrm{mix}}\rangle](E)$ and $\langle\widetilde{\rho}_{\mu,\,1000}^{\ \textrm{mix}}\rangle(E) - \sigma[\langle\widetilde{\rho}_{\mu,\,1000}^{\ \textrm{mix}}\rangle](E)$ are linearly interpolated in the energy range under consideration. 
Finally, the uncertainties on $\langle\rho_{\mu,\,1000}^{\textrm{mix}}/E_R\rangle(E_{Ri})$ are obtained by inserting each interpolated function in Eq.~(\ref{EqRhoAv}) and performing the integrals.

\section{Results} 
\label{secResults}

\begin{figure}[!ht]
\includegraphics[width=\linewidth]{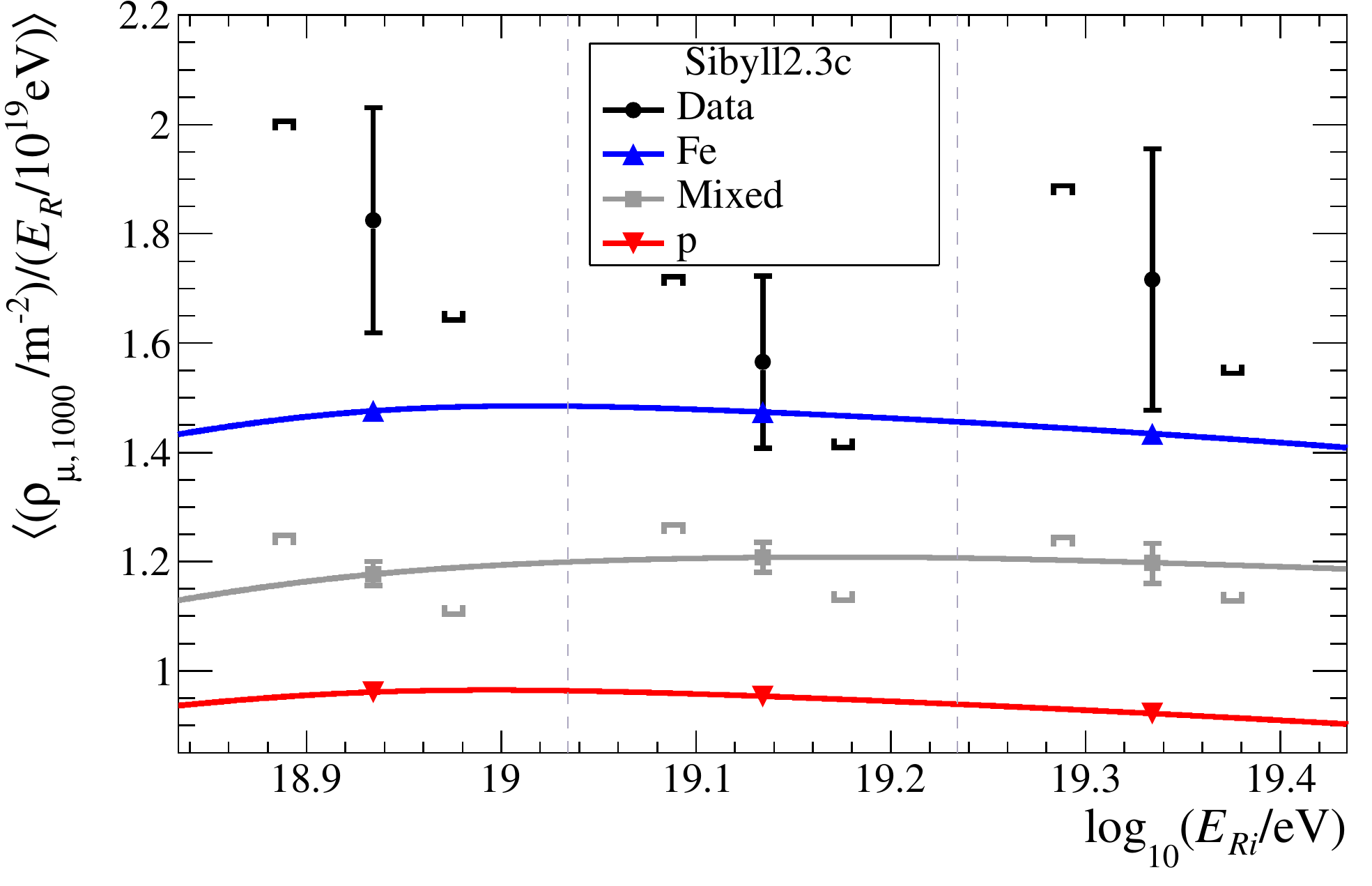}
\includegraphics[width=\linewidth]{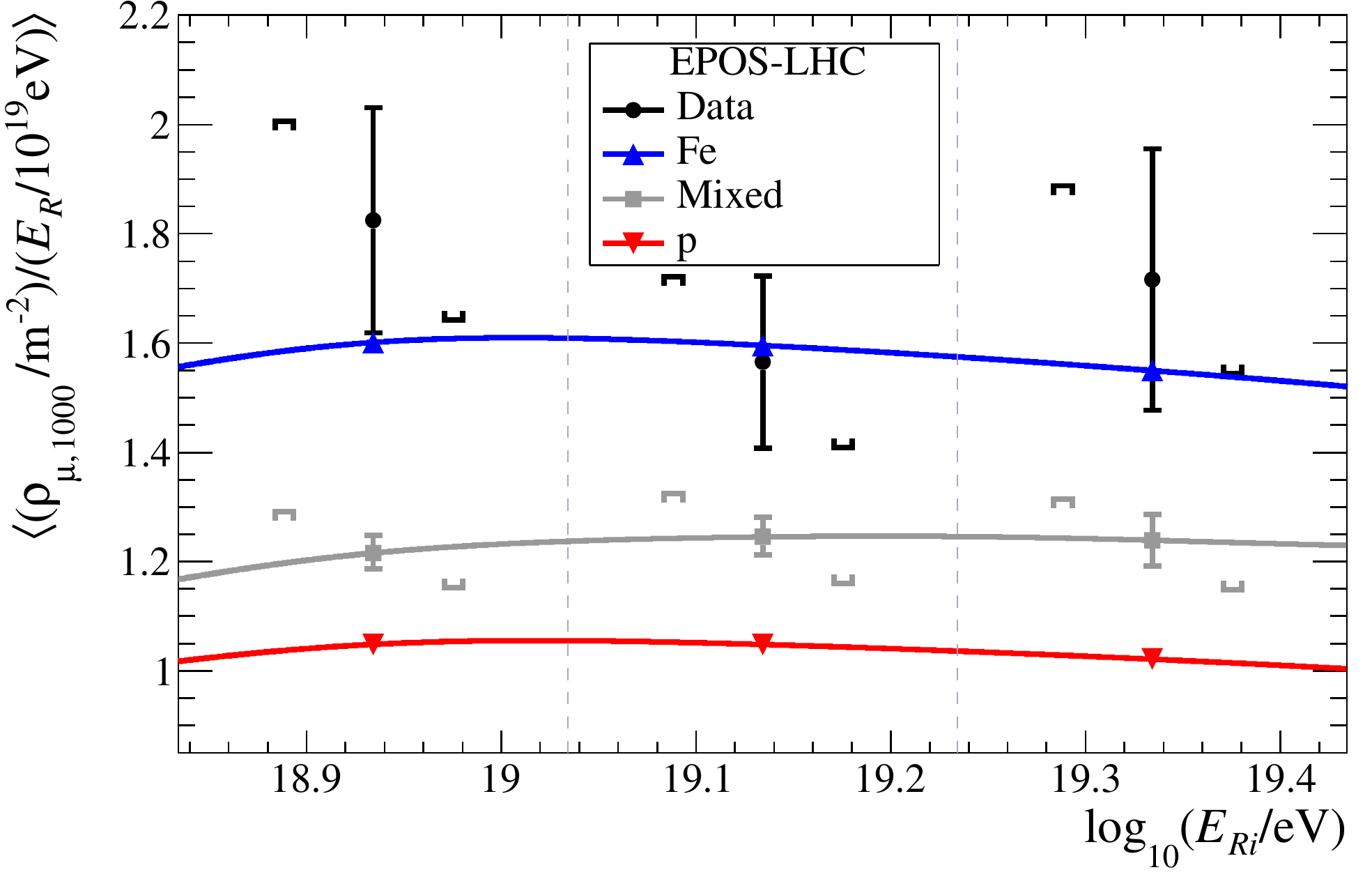}
\includegraphics[width=\linewidth]{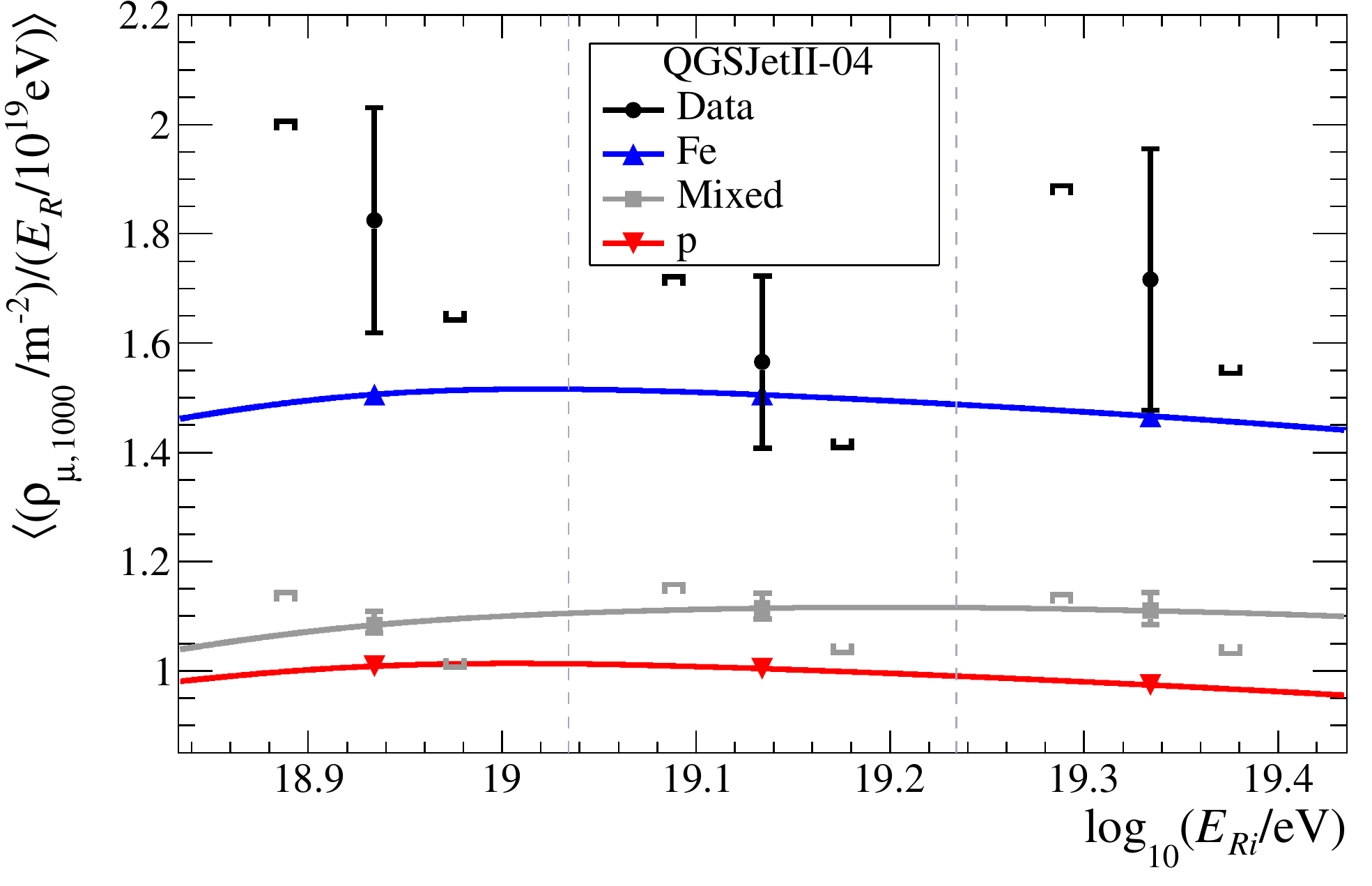}
\caption{Average muon density divided by the reconstructed energy, as a function of the logarithm of the reconstructed energy in the center of the $i$-th bin. Superimposed to AGASA data points \cite{Shinozaki2004} are the predictions for proton (red) and iron (blue) primaries, 
and for the mixed composition scenario corresponding to the models Sibyll2.3c (top panel), EPOS-LHC (middle panel), and QGSJetII-04 
(lower panel). The systematic uncertainties are enclosed by square brackets. The latter account for systematic uncertainties in the 
energy scale, hence they are diagonal, and also in the mass fractions in the case of the mixed composition scenarios. The vertical dashed lines correspond to the limits of the reconstructed energy bins considered. \label{fRhoMuOverEr}}
\end{figure}

Figure \ref{fRhoMuOverEr} shows $\langle\rho_{\mu,\,1000}/E_R\rangle(E_{Ri})$ as a function of the logarithm of the reconstructed energy bin obtained for AGASA data, proton and iron simulations, and for the mixed composition scenarios. The three data points represent the average of 
$67$, $33$ and $20$ events (from lower to higher energy), which correspond to a total of $120$ events. The square brackets associated 
to the AGASA data represent the systematic uncertainties corresponding to the reference energy scale. The square brackets associated to the 
mixed composition scenarios represent also the systematic uncertainties, which include the ones corresponding to the mass fractions 
and the one corresponding to the reference energy scale. Note that the latter are the dominant in this case. The bin width used in this analysis is $\Delta\log_{10}(E/\textrm{eV})=0.2$. It can be seen from Fig.~\ref{fRhoMuOverEr} that for all models, AGASA data points are incompatible with the mixed composition scenarios, with the only exception given by the middle bin 
corresponding to EPOS-LHC, when the reference energy scale is shifted to the right in $10\,\%$. However, the AGASA data are compatible 
with iron nuclei.

\begin{figure}[!ht]
\includegraphics[width=\linewidth]{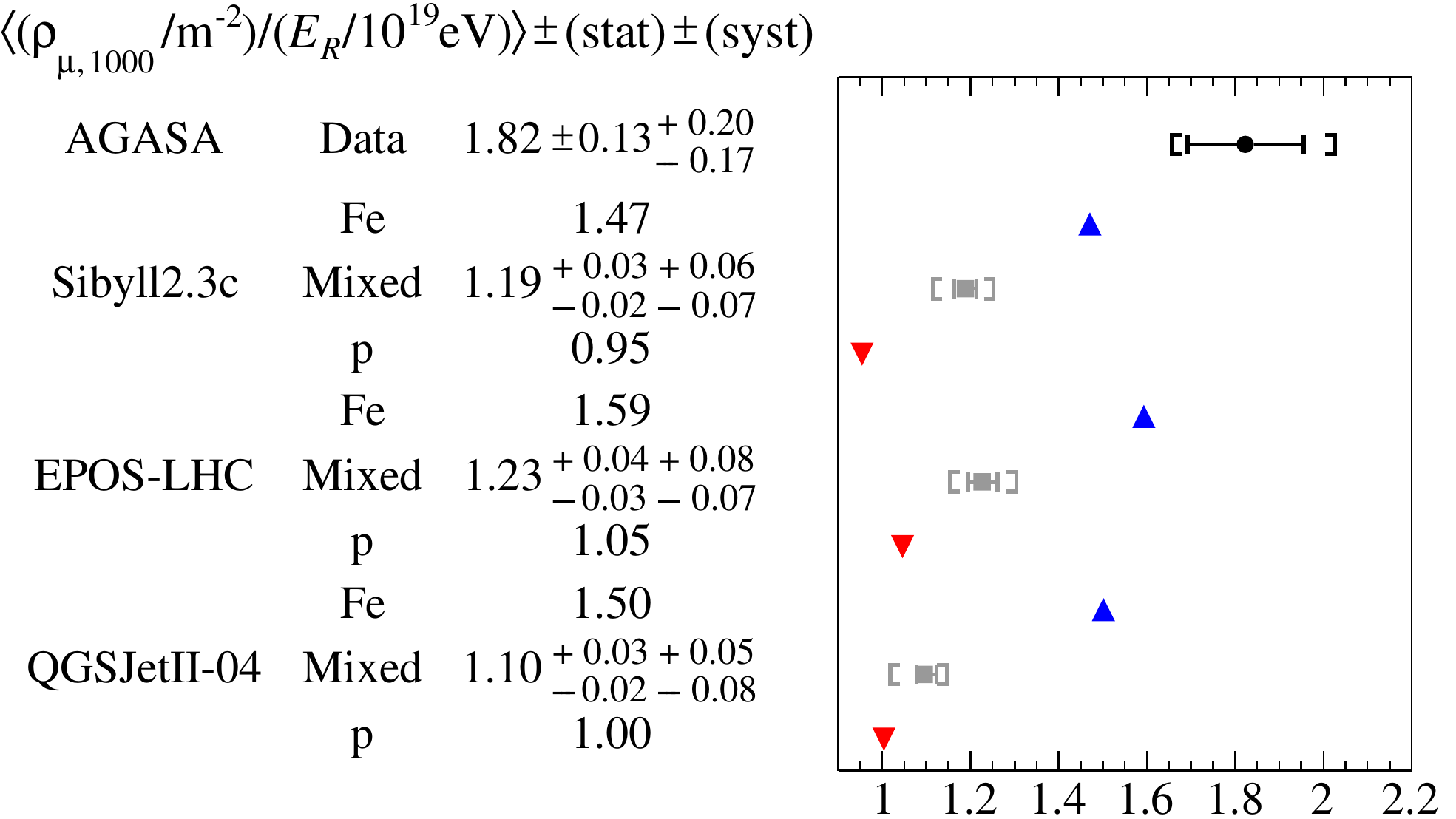}
\caption{Average muon density divided by the reconstructed energy for AGASA data and for proton, iron, and mixed composition scenarios. The obtained values are reported in the table (left) and are also plotted (right) on the same line.
The energy range under consideration is $18.83\,\leq\,\log_{10}(E_{R}/\textrm{eV})\,\leq\,19.46$. The analysed models are 
QGSJetII-04, EPOS-LHC, and Sibyll2.3c. The square brackets correspond to the systematic uncertainties.\label{fThreeBin}} 
\end{figure}

To further study the compatibility between AGASA data and the predictions corresponding to single primaries and mixed composition scenarios, a single value for $\langle\rho_{\mu,\,1000} / E_R\rangle$ is calculated taking the average in the energy range $18.83\,\leq\,\log_{10}(E_{R}/\textrm{eV})\,\leq\,19.46$. This is obtained for AGASA data and for the scenarios mentioned before. It is reasonable to compute such average over
the whole analysed energy range as $\rho_{\mu,\,1000} / E_R$ is nearly constant within this range. In 
Fig.~\ref{fThreeBin} the values of $\langle\rho_{\mu,\,1000} / E_R\rangle$ estimated from AGASA measurements, and the ones corresponding to
proton, iron, and mixed composition scenarios obtained by using the three models considered are shown.

As in Fig.~\ref{fRhoMuOverEr}, from Fig.~\ref{fThreeBin} it can also be seen that the composition inferred from $\langle\rho_{\mu,\,1000} / E_R\rangle$ obtained from AGASA data is compatible with heavy primaries, for the three models considered. This interpretation is inconsistent with the mixed composition scenarios. The discrepancies can be quantified in \enquote{sigmas}, considering the total low uncertainty for the AGASA data point and the total high uncertainties for the mixed composition scenarios. The resulting discrepancies are: $2.6\,\sigma$ for EPOS-LHC, $2.9\,\sigma$ for Sibyll2.3c, and $3.3\,\sigma$ for QGSJetII-04. 

As shown before, the composition of UHECRs inferred from the muon content of the showers detected by AGASA is incompatible with 
the one obtained from the $X_{\textrm{max}}$ measurements, when current interaction models are used to simulate the air showers required to 
interpret the data. It can be assumed that the composition (mass fractions) derived from the $X_{\textrm{max}}$ parameter is subject to 
smaller systematic uncertainties introduced by the models. Therefore, the discrepancies between $\langle\rho_{\mu,\,1000} / E_R\rangle$ obtained 
from AGASA data and the one corresponding to the mixed composition scenarios can be explained in terms of a muon deficit in air 
shower simulations. 

\begin{figure}[!ht]
\includegraphics[width=\linewidth]{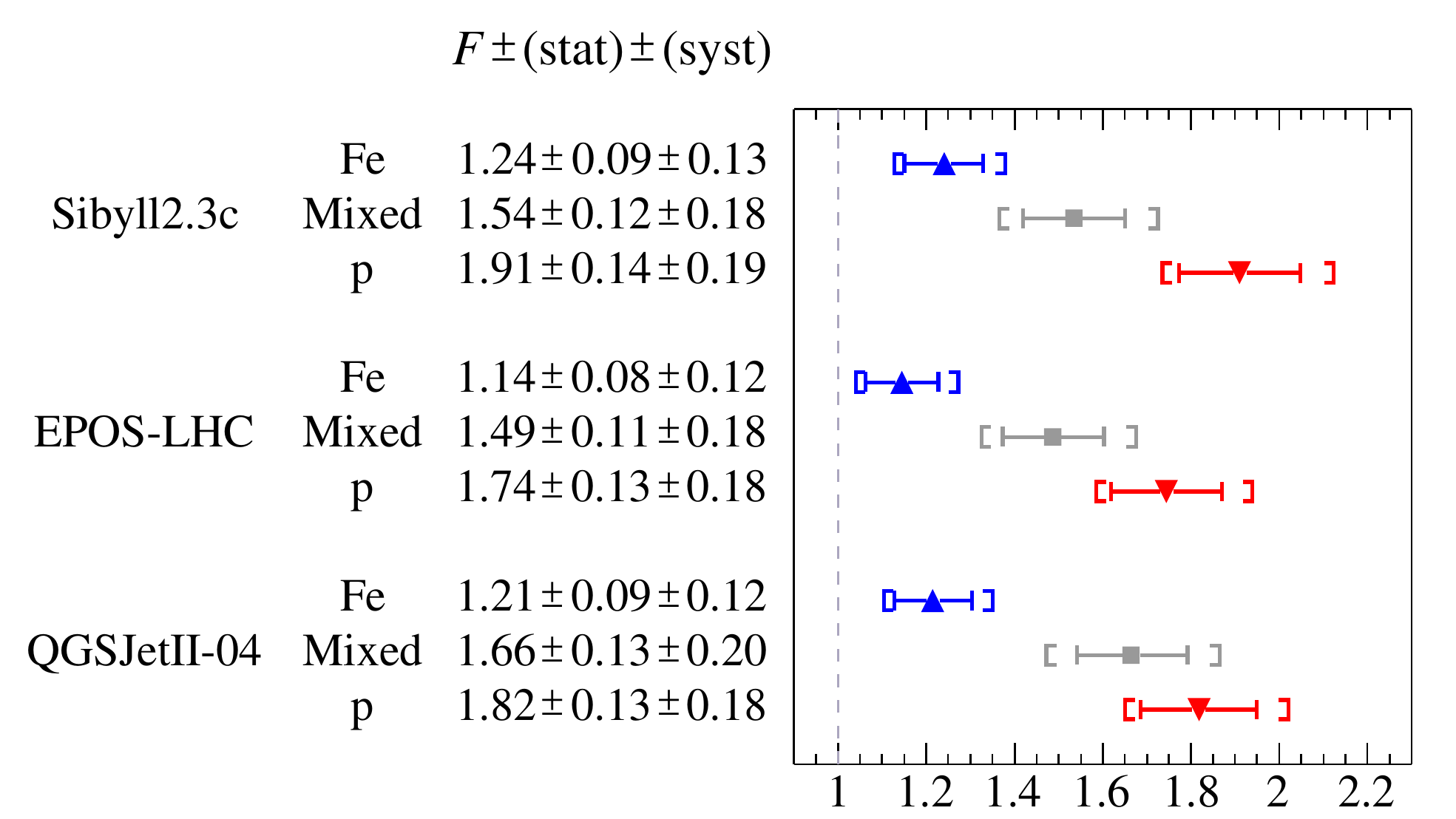}
\caption{Muon density correction factor $F$ corresponding to the single nuclei and mixed composition scenarios. The obtained values are reported in the table (left) and are also plotted (right) on the same line. The analysed models are QGSJetII-04, EPOS-LHC, and Sibyll2.3c. The energy range under consideration is $18.83\,\leq\,\log_{10}(E_{R}/\textrm{eV})\,\leq\,19.46$. The square brackets correspond to the systematic uncertainties. \label{fconfint}}
\end{figure}

The average muon deficit in the reconstructed energy range $18.83\,\leq\,\log_{10}(E_{R}/\textrm{eV})\,\leq\,19.46$ can be quantified by 
a correction factor $F$ which is defined as the ratio between the experimental average muon density divided by the energy and the one 
obtained from air shower simulations,
\begin{equation}
\label{eqF}
F = \ddfrac{\langle\rho_{\mu,\,1000}^{\textrm{data}} / E_R\rangle}{\langle\rho_{\mu,\,1000}^{\textrm{S}} / E_R\rangle},
\end{equation}
where S denotes the scenario under analysis, i.e.~S$=$\{mix, p, Fe\}. The uncertainties in $F$ are derived by propagating the uncertainties of $\langle\rho_{\mu,\,1000}^{\textrm{data}} / E_R\rangle$ and $\langle\rho_{\mu,\,1000}^{\textrm{S}} / E_R\rangle$. The obtained values of the correction factor $F$ with their statistic and systematic uncertainties, for the three models considered and for the single nuclei and mixed composition scenarios, are shown in Fig.~\ref{fconfint}.   

As mentioned in Section \ref{sec2}, the systematic uncertainty of the reference energy scale is taken as $10\,\%$. In a more conservative approach, the most extreme boundaries set by Auger and Telescope Array could be taken instead. This would lead to a systematic uncertainty in energy of $^{+27}_{-18}\,\%$. In this case, the correction factors $F$ for the mixed composition scenarios take the following values: $1.49 \pm 0.11\,\textrm{(stat)} \pm 0.34\,\textrm{(syst)}$ for EPOS-LHC, $1.54\pm 0.12\,\textrm{(stat)} \pm 0.35\,\textrm{(syst)}$ 
for Sibyll2.3c, and $1.66 \pm 0.13\,\textrm{(stat)} \pm 0.38\,\textrm{(syst)}$ for QGSJetII-04. It is remarkable that even in the most conservative approach, the models are not compatible with AGASA measurements within total uncertainties.

Moreover, the results presented in Figs.~\ref{fThreeBin} and \ref{fconfint} are essentially independent of the chosen flux 
parameterization. If the fit to the flux measurements of Auger \cite{Fenu17} are used instead of that of Telescope Array 
(see Appendix \ref{secAppFunctions}), the values of $\langle\rho_{\mu,\,1000}^{\textrm{S}} / E_R\rangle$ and $F$ change in less than $\sim 1\,\%$.

The muon deficit found in this analysis is qualitatively compatible with those obtained by the Pierre Auger \cite{MuonDeficit16} and Telescope Array Collaborations \cite{TAmuons}. However, a quantitative comparison with their results is not appropriate, since the studied phase-spaces (energy, zenith angle, distance to the shower axis) and variables under analysis differ.
 
It is worth mentioning that, with the surface scintilliator detectors of the upgrade of the Pierre Auger Observatory \cite{AugerPrime:16}, AugerPrime, it will be possible to study the muon deficit in air shower simulations in much more detail in the energy range considered in this work.

\section{Conclusions} 
\label{secConc}

The measurements of the muon density at $1000\,\textrm{m}$ from the shower axis obtained by the AGASA experiment have been analysed 
and compared to the predictions corresponding to single proton and iron primaries, as well as four-component mixed composition scenarios,
which are based on the $X_{\textrm{max}}$ measurements performed by Auger. The data analysis has been performed by using air shower
simulations generated with the high-energy hadronic interaction models QGSJetII-04, EPOS-LHC, and Sibyll2.3c. Furthermore, the reference
energy scale introduced by the \textit{Spectrum Working Group} \cite{Ivanov17} has been used in the performed analysis. Biases introduced by binning in energy and by a broad resolution in the energy reconstruction have been taken into account.

The AGASA measurements are found to be compatible with iron primaries for the interaction models used in the analyses. However, the AGASA muon measurements are incompatible with the predictions corresponding to the mixed composition scenarios for all models considered, in $2.6\,\sigma$ for EPOS-LHC, $2.9\,\sigma$ for Sibyll2.3c, and $3.3\,\sigma$ for QGSJetII-04. The discrepancies are larger if the energy scale is decreased. A $10\,\%$ systematic uncertainty in the energy scale was assumed. Nevertheless, the inconsistency between the mixed composition scenarios and AGASA data remains even when more conservative systematic uncertaintes in the energy scale are considered. 

Interpreting this incompatibility as a muon deficit in simulated air showers, a uniform muon density correction factor in the energy range $18.83\,\leq\,\log_{10}(E_{R}/\textrm{eV})\,\leq\,19.46$ was estimated for the interaction models considered. 
Therefore, for the mixed composition scenarios to be compatible with AGASA measurements, the muon density should be incremented by a factor of $1.49 \pm 0.11\,\textrm{(stat)} \pm 0.18\,\textrm{(syst)}$ for EPOS-LHC, $1.54 \pm 0.12\,\textrm{(stat)} \pm 0.18\,\textrm{(syst)}$ for Sibyll2.3c, and $1.66\pm0.13\,\textrm{(stat)}\pm0.20\,\textrm{(syst)}$ for QGSJetII-04. It is worth mentioning that the estimated muon deficits are qualitatively in agreement with the ones reported by the Pierre Auger and Telescope Array Collaborations.

\appendix

\section{Functions intervening in the calculation of $\langle \rho_{\mu,\,1000}/E_R \rangle$}
\label{secAppFunctions}

As mentioned in Section \ref{sec3}, the simulated average muon density at $1000\,\textrm{m}$ from the shower axis is fitted using a power law 
in energy (see Eq.~(\ref{eqLinFit})), for every primary and interaction model under consideration. The fits are performed within the range $18.0\,<\,\log_{10}(E/\textrm{eV}) \,<\,19.8$. The parameters obtained as a result of the fits are reported in Table \ref{TableFits}.
\begin{table}[ht]
\begin{ruledtabular}
\begin{tabular}{cccc}
Primary & Model & $\rho_{\mu (19)} \ [\textrm{m}^{-2}]$ & $\beta$ \\[0.1cm]
\hline \\[-0.2cm]
\multirow{3}{*}{p} & QGSJetII-04 &  $1.203\pm 0.011$ & $0.887\pm 0.007$ \\
                   & EPOS-LHC    &  $1.253\pm 0.013$ & $0.897\pm 0.007$ \\
                   & Sibyll2.3c  &  $1.145\pm 0.014$ & $0.880\pm 0.009$ \\ [0.2cm]
\multirow{3}{*}{He} & QGSJetII-04 &  $1.367\pm 0.009$ & $0.905\pm 0.005$ \\
                   & EPOS-LHC    &  $1.422\pm 0.011$ & $0.897\pm 0.006$ \\
                   & Sibyll2.3c  & $1.309\pm 0.010$ & $0.900\pm 0.006$ \\[0.2cm]
\multirow{3}{*}{N} & QGSJetII-04 &  $1.555\pm 0.009$ & $0.892\pm 0.004$ \\
                   & EPOS-LHC    &  $1.634\pm 0.008$ & $0.894\pm 0.004$ \\
                   & Sibyll2.3c  & $1.509\pm 0.008$ & $0.890\pm 0.004$ \\[0.2cm]
\multirow{3}{*}{Fe} & QGSJetII-04 &  $1.800\pm 0.005$ & $0.896\pm 0.002$ \\
                   & EPOS-LHC    &  $1.911\pm 0.006$ & $0.890\pm 0.002$ \\
                   & Sibyll2.3c  & $1.762\pm 0.007$ & $0.894\pm 0.002$ \\[0.1cm]
\end{tabular}
\end{ruledtabular}
\caption{Fitted values of the parameters corresponding to $\langle\widetilde{\rho}_{\mu,\,1000} \rangle(E) = \rho_{\mu (19)} \left( E/10^{19}\,\textrm{eV} \right)^{\beta}$ (Eq.~(\ref{eqLinFit})), for proton, helium, nitrogen, and iron primaries, for the models QGSJetII-04, EPOS-LHC, and Sibyll2.3c.}
\label{TableFits}
\end{table}

\begin{figure}[!ht]
\includegraphics[width=\linewidth]{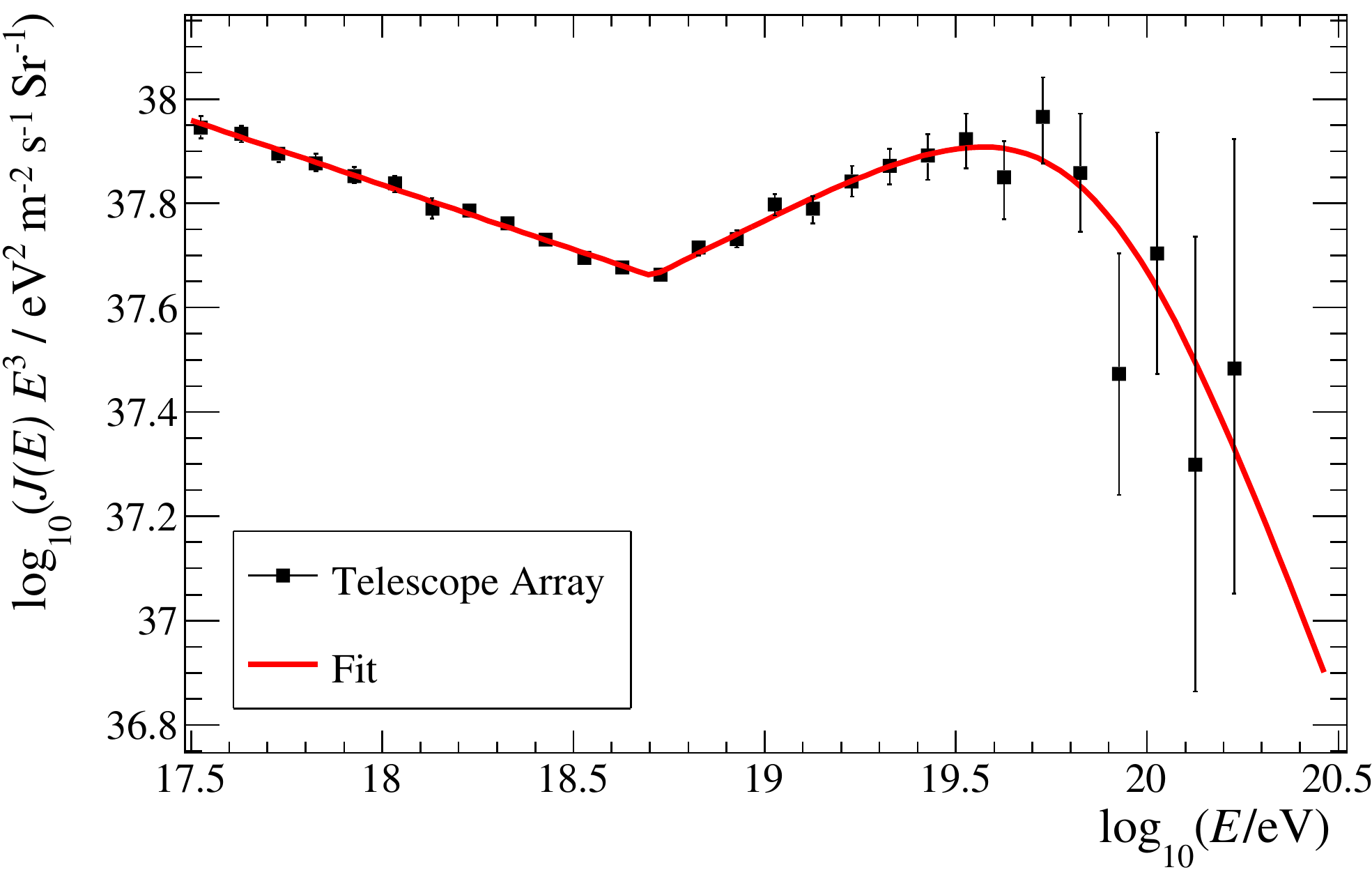}
\caption{Logarithm of the UHECR flux multiplied by the energy to the power of three as a function of the logarithm of the energy. The
data points correspond to the measurments done by Telescope Array \cite{TAFlux} and the solid line corresponds to the fit of the data 
(see text for details). The energy scale of Telescope Array is shifted to the reference energy scale. \label{fflux}}
\end{figure}

The UHECR flux measured by Telescope Array, shifted to the reference energy scale as explained in Section \ref{sec2}, is fitted using 
the following function \cite{Fenu17},
\begin{equation}
    J(E) =  A \left\{ 
                \begin{array}{ll}
                \! \! \! \!\left(\ddfrac{E}{E_a} \right)^{-\gamma_1} &  \log E \leq \log E_a \\ [0.2cm]
                \! \!\! \! \left( \ddfrac{E}{E_a} \right)^{-\gamma_2} \ddfrac{1+\left( E_a/E_s \right)^{\delta\gamma}}{1+\left(E/E_s\right)^{\delta\gamma}} &  \log E > \log E_a
                \end{array}
            \right. \! \! ,
\label{eqflux}
\end{equation}
where $A$, $E_a$, $E_s$, $\gamma_1$, $\gamma_2$, and $\delta\gamma$ are free fit parameters. Fig.~\ref{fflux} shows the fit of the 
Telescope Array data. The resulting values of the parameters are given in Table \ref{TableApp}.
\begin{table}[!ht]
\begin{ruledtabular}
\begin{tabular}{lc}
 Parameter &  Fitted Value \\
\hline \\[-0.2cm]
 $A$ $[10^{-19} \textrm{eV} \textrm{km}^{2} \ \textrm{yr sr}]$  & $3.5 \pm 0.5$  \\ %2.9
 $\log_{10} (E_a/\textrm{eV})$  & $18.71 \pm 0.02$ \\ %18.73
 $\log_{10} (E_s/\textrm{eV})$  & $19.88 \pm 0.09$ \\ %19.90
 $\gamma_1$       & $3.248 \pm 0.012$ \\  
 $\gamma_2$       & $2.63 \pm 0.06$ \\ 
 $\delta\gamma$       & $2.4 \pm 0.8$ \\[0.1cm]  
\end{tabular}
\end{ruledtabular}
\caption{Parameters of the fit to the UHECR flux measured by Telescope Array (see Eq.~(\ref{eqflux})).}
\label{TableApp}
\end{table}

As mentioned in Section \ref{sec1}, the conditional distribution function of the AGASA reconstructed energy $E_R$, given the 
\enquote{true} energy $E$, follows a log-normal distribution \cite{Takeda2003},
\begin{equation}
G(E_R|E) = \ddfrac{1}{\sqrt{2 \pi}\ \sigma(E) E_R } \ \exp \left[-\ddfrac{\ln^2(E_R/E)}{2 \sigma^2(E)}\right],  \\
\label{eqG}
\end{equation}
where the parameter $\sigma$ is related to the variance of $E_R$ through,
\begin{eqnarray}
\sigma_R^2(E) &=& \left( \exp \left[\sigma^2(E) \right]-1 \right) \times \nonumber \\
&&\exp \left[2 \ln (E/\textrm{eV})+\sigma^2(E) \right].
\label{esigmas1}
\end{eqnarray}

From Eq.~(\ref{esigmas1}) it is possible to obtain the parameter $\sigma$ of the log-normal distribution as a function of 
$\sigma_R$,
\begin{equation}
\sigma(E) = \sqrt{ \ln\left[ \frac{1}{2}+\frac{1}{2} \sqrt{1+4 \, \frac{\sigma_R^2(E)}{E^2}} \right]}.
\label{esigmas2} 
\end{equation}
Therefore, $G(E_R|E)$ is completely determined providing the function $\sigma_R(E)$.   

$\sigma_R(E)$ is obtained from the signal resolution $\sigma[S_{600}]$ as a function of $\log_{10} S_{600} $ by using the 
$S_{600}$ to energy conversion function, reported in Ref.~\cite{Takeda2003}, corrected to match the reference energy scale 
as explained in Section \ref{sec2}, i.e.~$E = 0.68 \times 2.21 \times 10^{17}\ S_{0}(600)^{1.03}\, \textrm{eV}$. The $S_{600}$ 
resolution as a function of $\log_{10} (S_{600})$, obtained from shower and detector simulations, for showers with zenith 
angles in $33\degree \leq \theta \leq 44\degree$, is taken from Ref.~\cite{Yoshida95}.  

\begin{figure}[hb]
\includegraphics[width=\linewidth]{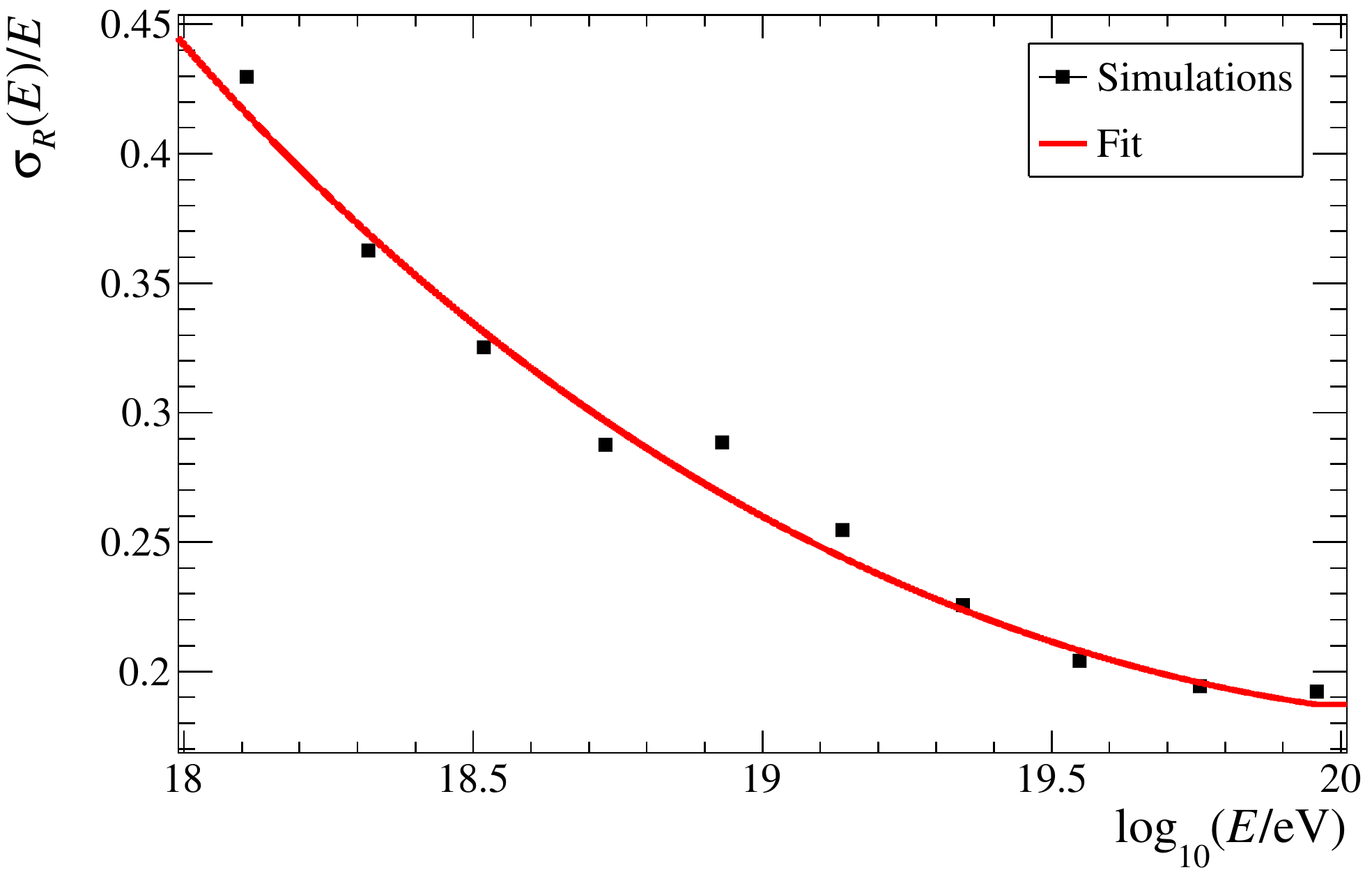}
\caption{Relative reconstructed energy uncertainty as a function of the logarithm of the energy. The data points are obtained from 
simulations \cite{Yoshida95}. The zenith angles of the showers are in the range $33\degree \leq \theta \leq 44\degree$. The solid line corresponds to an approximating function (see the text for details). \label{fepsilon}}
\end{figure}

%\vspace{-1mm}
Figure \ref{fepsilon} shows the relative reconstructed energy uncertainty as a function of the logarithm of the energy. It is 
a decreasing function of energy since higher energy events are reconstructed with smaller uncertainties. The data points 
corresponding to $\sigma(E)$ are calculated from the ones corresponding to $\sigma_R(E)$ and using Eq.~(\ref{esigmas2}). The resulting values for $\sigma(E)$ are then fitted using a second degree polynomial in $\log_{10} (E/\textrm{eV})$ given by
%  
%\vspace{-1mm}
\begin{eqnarray}
\sigma(E) &=& (17 \pm 3) - (1.59 \pm 0.37) \log_{10} (E_R/\textrm{eV}) \nonumber  \\
          && + (0.039 \pm 0.009) \log_{10}^{2} (E_R/\textrm{eV}).
\label{esigmas3}    
\end{eqnarray}

Finally, $\sigma_R(E)/E$ is obtained by using the expression for $\sigma(E)$ given by Eq.~(\ref{esigmas3}) and Eq.~(\ref{esigmas1}). The function for $\sigma_R(E)/E$ obtained in this way is shown in Fig.~\ref{fepsilon} (solid line). It can be seen that it is in very good agreement 
with the $\sigma_R(E)/E$ data points.

\section{AGASA data set}
\label{secAppRawData}

The AGASA measurements that are used in this work are reported in Table \ref{TableRawData}. This data set is extracted from Fig.~7 of Ref.~\cite{Shinozaki2004}. The reconstructed energy of the events listed in Table \ref{TableRawData} is expressed in the reference energy
scale (see Sec.~\ref{sec2}).

\begin{table*}[!t]
\medskip
\setlength\tabcolsep{4pt}
\begin{ruledtabular}
\begin{tabular}{@{}*{2}{c} >{\color{white}\vrule width12pt}c*{2}{c}@{}}
 $\log_{10} (E/\textrm{eV})$ &  $\log_{10} (\rho_{\mu,\,1000}/\textrm{m}^{2})$ & & $\log_{10} (E/\textrm{eV})$ &  $\log_{10} (\rho_{\mu,\,1000}/\textrm{m}^{2})$ \\
\cline{1-2}\cline{4-5}
  18.835 & -0.003   & &  18.988 & 0.303 \\ 
  18.835 & 0.031   & &   18.989 & -0.344 \\ 
  18.836 & $-\infty$   & &  18.995 & $-\infty$ \\
  18.837 & 0.190   & &   19.010 & 0.129 \\ 
  18.847 & -0.081   & &  19.019 & 0.320 \\ 
  18.849 & 0.128   & &   19.021 & 0.112 \\ 
  18.849 & -0.092   & &  19.027 & -0.047 \\ 
  18.853 & 0.333   & &   19.035 & 0.116 \\ 
  18.853 & 0.506   & &   19.035 & -0.120 \\ 
  18.853 & -0.127   & &  19.037 & 0.228 \\ 
  18.856 & 0.009   & &   19.038 & -0.088 \\ 
  18.859 & 0.581   & &   19.039 & 0.112 \\ 
  18.862 & 0.037   & &   19.041 & 0.170 \\ 
  18.862 & 0.805   & &   19.056 & -0.150 \\ 
  18.864 & -0.016   & &  19.061 & 0.028 \\ 
  18.864 & -0.147   & &  19.069 & 0.533 \\ 
  18.865 & 0.100   & &   19.075 & 0.333 \\ 
  18.868 & -0.108   & &  19.075 & 0.329 \\ 
  18.870 & 0.046   & &   19.077 & 0.648 \\ 
  18.873 & 0.245   & &   19.080 & 0.111 \\ 
  18.874 & -0.298   & &  19.084 & 0.471 \\ 
  18.875 & 0.311   & &   19.085 & $-\infty$ \\
  18.877 & 0.024   & &   19.092 & 0.530 \\ 
  18.882 & 0.317   & &   19.095 & 0.643 \\ 
  18.886 & 0.353   & &   19.100 & 0.231 \\ 
  18.889 & 0.411   & &   19.101 & 0.310 \\ 
  18.889 & 0.441   & &   19.106 & 0.367 \\ 
  18.893 & -0.433   & &  19.114 & 0.535 \\ 
  18.894 & -1.446   & &  19.116 & -0.057 \\ 
  18.895 & -0.392   & &  19.118 & 0.396 \\ 
  18.900 & 0.163   & &   19.120 & 0.529 \\ 
  18.900 & -0.208   & &  19.127 & 0.276 \\ 
  18.903 & 0.038   & &   19.147 & 0.652 \\ 
  18.905 & 0.207   & &   19.152 & -0.127 \\ 
  18.906 & 0.560   & &   19.165 & 0.586 \\ 
  18.907 & 0.041   & &   19.172 & 0.111 \\ 
  18.908 & -0.496   & &  19.185 & 0.338 \\ 
  18.915 & 0.187   & &   19.189 & 0.398 \\ 
  18.917 & 0.058   & &   19.211 & -0.231 \\ 
  18.919 & 0.080   & &   19.233 & 0.516 \\ 
  18.921 & -0.409   & &  19.237 & 0.507 \\ 
  18.923 & -0.255   & &  19.246 & $-\infty$ \\
  18.927 & $-\infty$   & &  19.257 & 0.388 \\ 
  18.935 & 0.157   & &   19.265 & 0.581 \\ 
  18.936 & -0.976   & &  19.286 & 0.068 \\ 
  18.938 & 0.442   & &   19.288 & 0.648 \\ 
  18.938 & 0.272   & &   19.288 & 0.490 \\ 
  18.944 & 0.759   & &   19.290 & 0.947 \\ 
  18.945 & 0.153   & &   19.290 & 0.732 \\ 
  18.948 & 0.352   & &   19.296 & 0.789 \\ 
  18.955 & -0.466   & &  19.299 & 0.405 \\ 
  18.955 & -0.218   & &  19.301 & -0.080 \\ 
  18.959 & 0.301   & &   19.303 & 0.226 \\ 
  18.964 & 0.159   & &   19.335 & 0.555 \\ 
  18.965 & 0.343   & &   19.350 & 0.485 \\ 
  18.965 & 0.311   & &   19.353 & 0.603 \\ 
  18.966 & -0.130   & &  19.360 & 0.700 \\ 
  18.978 & 0.019   & &   19.363 & 0.035 \\ 
  18.979 & 0.349   & &   19.387 & 0.810 \\ 
  18.987 & 0.923   & &   19.424 & 0.860 \\  [0.1cm]
\end{tabular}
\end{ruledtabular}
\caption{Logarithm of the reconstructed energy in the reference energy scale (see Sec.~\ref{sec2}) and logarithm of the muon density at $1000\,\text{m}$ from the shower axis, for each of the events considered in this work. These values are extracted from Fig.~7 of Ref.~\cite{Shinozaki2004}.}
\label{TableRawData}
\end{table*}
\section{Parameters of the simulations}\label{secAppParams}

Following Ref.~\cite{Takeda2003}, the altitude used for the simulations is the average altitude of the detectors, $667\,\textrm{m}$. 
The $x$ and $z$ components of the Earth's magnetic field, in the CORSIKA coordinate system \cite{CORSIKA}, at Akeno, Yamanashi 
(Hokuto, Yamanashi since 2004) are set to $B_x=\SI{30.13}{\micro\tesla}$ and $B_z=\SI{35.45}{\micro\tesla}$ \cite{Geomag}. In 
order to speed up the simulations the thinning algorithm implemented in the CORSIKA program is used \cite{Kobal2001}. A thinning 
level of $10^{-6}$ with a maximum weight factor given by $10^{-6}\times (E/\textrm{GeV})$ is considered, where $E$ is the input energy of
the incident cosmic ray (see Ref.~\cite{CORSIKA} for details).

\begin{acknowledgments}

A.D.S.~and A.E.~are members of the Carrera del Investigador Cient\'ifico of CONICET, Argentina. The authors acknowledge the anonymous 
reviewer and the members of the Pierre Auger Collaboration, specially T. Pierog, D. Ravignani, and R. Clay. This work is supported by 
ANPCyT PICT-2015-2752.

\end{acknowledgments}

\clearpage
\bibliography{MyBibliography}{}

%merlin.mbs apsrev4-1.bst 2010-07-25 4.21a (PWD, AO, DPC) hacked
%Control: key (0)
%Control: author (8) initials jnrlst
%Control: editor formatted (1) identically to author
%Control: production of article title (-1) disabled
%Control: page (0) single
%Control: year (1) truncated
%Control: production of eprint (0) enabled
\begin{thebibliography}{33}%
\makeatletter
\providecommand \@ifxundefined [1]{%
 \@ifx{#1\undefined}
}%
\providecommand \@ifnum [1]{%
 \ifnum #1\expandafter \@firstoftwo
 \else \expandafter \@secondoftwo
 \fi
}%
\providecommand \@ifx [1]{%
 \ifx #1\expandafter \@firstoftwo
 \else \expandafter \@secondoftwo
 \fi
}%
\providecommand \natexlab [1]{#1}%
\providecommand \enquote  [1]{``#1''}%
\providecommand \bibnamefont  [1]{#1}%
\providecommand \bibfnamefont [1]{#1}%
\providecommand \citenamefont [1]{#1}%
\providecommand \href@noop [0]{\@secondoftwo}%
\providecommand \href [0]{\begingroup \@sanitize@url \@href}%
\providecommand \@href[1]{\@@startlink{#1}\@@href}%
\providecommand \@@href[1]{\endgroup#1\@@endlink}%
\providecommand \@sanitize@url [0]{\catcode `\\12\catcode `\$12\catcode
  `\&12\catcode `\#12\catcode `\^12\catcode `\_12\catcode `\%12\relax}%
\providecommand \@@startlink[1]{}%
\providecommand \@@endlink[0]{}%
\providecommand \url  [0]{\begingroup\@sanitize@url \@url }%
\providecommand \@url [1]{\endgroup\@href {#1}{\urlprefix }}%
\providecommand \urlprefix  [0]{URL }%
\providecommand \Eprint [0]{\href }%
\providecommand \doibase [0]{http://dx.doi.org/}%
\providecommand \selectlanguage [0]{\@gobble}%
\providecommand \bibinfo  [0]{\@secondoftwo}%
\providecommand \bibfield  [0]{\@secondoftwo}%
\providecommand \translation [1]{[#1]}%
\providecommand \BibitemOpen [0]{}%
\providecommand \bibitemStop [0]{}%
\providecommand \bibitemNoStop [0]{.\EOS\space}%
\providecommand \EOS [0]{\spacefactor3000\relax}%
\providecommand \BibitemShut  [1]{\csname bibitem#1\endcsname}%
\let\auto@bib@innerbib\@empty
%</preamble>
\bibitem [{\citenamefont {{S. Ostapchenko}}(2013)}]{QGSJet}%
  \BibitemOpen
  \bibfield  {author} {\bibinfo {author} {\bibnamefont {{S. Ostapchenko}}},\
  }\href {\doibase 10.1051/epjconf/20125202001} {\bibfield  {journal} {\bibinfo
   {journal} {EPJ Web of Conferences}\ }\textbf {\bibinfo {volume} {52}},\
  \bibinfo {pages} {02001} (\bibinfo {year} {2013})}\BibitemShut {NoStop}%
\bibitem [{\citenamefont {{T. Pierog}}\ \emph {et~al.}(2015)\citenamefont {{T.
  Pierog}}, \citenamefont {{Iu. Karpenko}}, \citenamefont {{J. M. Katzy}},
  \citenamefont {{E. Yatsenko}},\ and\ \citenamefont {{K. Werner}}}]{Epos}%
  \BibitemOpen
  \bibfield  {author} {\bibinfo {author} {\bibnamefont {{T. Pierog}}}, \bibinfo
  {author} {\bibnamefont {{Iu. Karpenko}}}, \bibinfo {author} {\bibnamefont
  {{J. M. Katzy}}}, \bibinfo {author} {\bibnamefont {{E. Yatsenko}}}, \ and\
  \bibinfo {author} {\bibnamefont {{K. Werner}}},\ }\href {\doibase
  10.1103/PhysRevC.92.034906} {\bibfield  {journal} {\bibinfo  {journal} {Phys.
  Rev.}\ }\textbf {\bibinfo {volume} {C92}},\ \bibinfo {pages} {034906}
  (\bibinfo {year} {2015})},\ \Eprint {http://arxiv.org/abs/1306.0121}
  {arXiv:1306.0121 [hep-ph]} \BibitemShut {NoStop}%
%%CITATION = ARXIV:1306.0121;%%
\bibitem [{\citenamefont {{F. Riehn}}\ \emph {et~al.}(2017)\citenamefont {{F.
  Riehn}}, \citenamefont {{H. P. Dembinski}}, \citenamefont {{R. Engel}},
  \citenamefont {{A. Fedynitch}}, \citenamefont {{T. K. Gaisser}},\ and\
  \citenamefont {{T. Stanev}}}]{Sibyll}%
  \BibitemOpen
  \bibfield  {author} {\bibinfo {author} {\bibnamefont {{F. Riehn}}}, \bibinfo
  {author} {\bibnamefont {{H. P. Dembinski}}}, \bibinfo {author} {\bibnamefont
  {{R. Engel}}}, \bibinfo {author} {\bibnamefont {{A. Fedynitch}}}, \bibinfo
  {author} {\bibnamefont {{T. K. Gaisser}}}, \ and\ \bibinfo {author}
  {\bibnamefont {{T. Stanev}}},\ }in\ \href@noop {} {\emph {\bibinfo
  {booktitle} {Proceedings of the 35th International Cosmic Ray Conference
  (ICRC)}}},\ Vol.\ \bibinfo {volume} {301}\ (\bibinfo {year} {2017})\ p.\
  \bibinfo {pages} {301},\ \Eprint {http://arxiv.org/abs/1709.07227}
  {arXiv:1709.07227 [hep-ph]} \BibitemShut {NoStop}%
\bibitem [{\citenamefont {Gaisser}\ \emph {et~al.}(2016)\citenamefont
  {Gaisser}, \citenamefont {Engel},\ and\ \citenamefont {Resconi}}]{Book}%
  \BibitemOpen
  \bibfield  {author} {\bibinfo {author} {\bibfnamefont {T.~K.}\ \bibnamefont
  {Gaisser}}, \bibinfo {author} {\bibfnamefont {R.}~\bibnamefont {Engel}}, \
  and\ \bibinfo {author} {\bibfnamefont {E.}~\bibnamefont {Resconi}},\ }\href
  {\doibase 10.1017/CBO9781139192194} {\emph {\bibinfo {title} {Cosmic Rays and
  Particle Physics}}},\ \bibinfo {edition} {2nd}\ ed.\ (\bibinfo  {publisher}
  {Cambridge University Press},\ \bibinfo {year} {2016})\BibitemShut {NoStop}%
\bibitem [{\citenamefont {{A. Aab}}\ \emph {et~al.}(2016)\citenamefont {{A.
  Aab}} \emph {et~al.}}]{AugerPrime:16}%
  \BibitemOpen
  \bibfield  {author} {\bibinfo {author} {\bibnamefont {{A. Aab}}} \emph
  {et~al.},\ }\href {\doibase 10.1051/epjconf/201920808003} {\bibfield
  {journal} {\bibinfo  {journal} {EPJ Web Conf.}\ }\textbf {\bibinfo {volume}
  {208}},\ \bibinfo {pages} {08003} (\bibinfo {year} {2016})},\ \Eprint
  {http://arxiv.org/abs/1604.03637} {arXiv:1604.03637} \BibitemShut {NoStop}%
\bibitem [{\citenamefont {{J. Bellido for the Pierre Auger
  Collaboration}}(2017)}]{Bellido18}%
  \BibitemOpen
  \bibfield  {author} {\bibinfo {author} {\bibnamefont {{J. Bellido for the
  Pierre Auger Collaboration}}},\ }in\ \href@noop {} {\emph {\bibinfo
  {booktitle} {Proceedings of the 35th International Cosmic Ray Conference
  (ICRC)}}},\ Vol.\ \bibinfo {volume} {301}\ (\bibinfo {year} {2017})\ p.\
  \bibinfo {pages} {506}\BibitemShut {NoStop}%
\bibitem [{\citenamefont {{T. Pierog}}(2017)}]{Pierog17}%
  \BibitemOpen
  \bibfield  {author} {\bibinfo {author} {\bibnamefont {{T. Pierog}}},\ }in\
  \href@noop {} {\emph {\bibinfo {booktitle} {Proceedings of the 35th
  International Cosmic Ray Conference (ICRC)}}},\ Vol.\ \bibinfo {volume}
  {301}\ (\bibinfo {year} {2017})\ p.\ \bibinfo {pages} {1100}\BibitemShut
  {NoStop}%
\bibitem [{\citenamefont {{R. R. Prado}}(2019)}]{Prado18}%
  \BibitemOpen
  \bibfield  {author} {\bibinfo {author} {\bibnamefont {{R. R. Prado}}},\
  }\href {\doibase 10.1051/epjconf/201920808003} {\bibfield  {journal}
  {\bibinfo  {journal} {EPJ Web Conf.}\ }\textbf {\bibinfo {volume} {208}},\
  \bibinfo {pages} {08003} (\bibinfo {year} {2019})}\BibitemShut {NoStop}%
\bibitem [{\citenamefont {{H. P. Dembinski \textit{et al.} for the EAS-MSU}}\
  \emph {et~al.}(2019)\citenamefont {{H. P. Dembinski \textit{et al.} for the
  EAS-MSU}}, \citenamefont {{IceCube}}, \citenamefont {{KASCADE-Grande}},
  \citenamefont {{NEVOD-DECOR}}, \citenamefont {{Pierre Auger}}, \citenamefont
  {{SUGAR}}, \citenamefont {{Telescope Array}},\ and\ \citenamefont {{Yakutsk
  EAS Array collaborations}}}]{Dembinski19}%
  \BibitemOpen
  \bibfield  {author} {\bibinfo {author} {\bibnamefont {{H. P. Dembinski
  \textit{et al.} for the EAS-MSU}}}, \bibinfo {author} {\bibnamefont
  {{IceCube}}}, \bibinfo {author} {\bibnamefont {{KASCADE-Grande}}}, \bibinfo
  {author} {\bibnamefont {{NEVOD-DECOR}}}, \bibinfo {author} {\bibnamefont
  {{Pierre Auger}}}, \bibinfo {author} {\bibnamefont {{SUGAR}}}, \bibinfo
  {author} {\bibnamefont {{Telescope Array}}}, \ and\ \bibinfo {author}
  {\bibnamefont {{Yakutsk EAS Array collaborations}}},\ }\href {\doibase
  10.1051/epjconf/201921002004} {\bibfield  {journal} {\bibinfo  {journal} {EPJ
  Web Conf.}\ }\textbf {\bibinfo {volume} {210}},\ \bibinfo {pages} {02004}
  (\bibinfo {year} {2019})}\BibitemShut {NoStop}%
\bibitem [{\citenamefont {{L. Cazon for the EAS-MSU}}\ \emph
  {et~al.}(2019)\citenamefont {{L. Cazon for the EAS-MSU}}, \citenamefont
  {{IceCube}}, \citenamefont {{KASCADE-Grande}}, \citenamefont {{NEVOD-DECOR}},
  \citenamefont {{Pierre Auger}}, \citenamefont {{SUGAR}}, \citenamefont
  {{Telescope Array}},\ and\ \citenamefont {{Yakutsk EAS Array
  collaborations}}}]{Cazon19likeDembinski19}%
  \BibitemOpen
  \bibfield  {author} {\bibinfo {author} {\bibnamefont {{L. Cazon for the
  EAS-MSU}}}, \bibinfo {author} {\bibnamefont {{IceCube}}}, \bibinfo {author}
  {\bibnamefont {{KASCADE-Grande}}}, \bibinfo {author} {\bibnamefont
  {{NEVOD-DECOR}}}, \bibinfo {author} {\bibnamefont {{Pierre Auger}}}, \bibinfo
  {author} {\bibnamefont {{SUGAR}}}, \bibinfo {author} {\bibnamefont
  {{Telescope Array}}}, \ and\ \bibinfo {author} {\bibnamefont {{Yakutsk EAS
  Array collaborations}}},\ }in\ \href@noop {} {\emph {\bibinfo {booktitle}
  {{Proceedings of the 36th International Cosmic Ray Conference (ICRC)}}}},\
  Vol.\ \bibinfo {volume} {358}\ (\bibinfo {year} {2019})\ p.\ \bibinfo {pages}
  {214}\BibitemShut {NoStop}%
\bibitem [{\citenamefont {Aab}\ \emph {et~al.}(2016)\citenamefont {Aab} \emph
  {et~al.}}]{MuonDeficit16}%
  \BibitemOpen
  \bibfield  {author} {\bibinfo {author} {\bibfnamefont {A.}~\bibnamefont
  {Aab}} \emph {et~al.} (\bibinfo {collaboration} {Pierre Auger
  Collaboration}),\ }\href {\doibase 10.1103/PhysRevLett.117.192001} {\bibfield
   {journal} {\bibinfo  {journal} {Phys. Rev. Lett.}\ }\textbf {\bibinfo
  {volume} {117}},\ \bibinfo {pages} {192001} (\bibinfo {year}
  {2016})}\BibitemShut {NoStop}%
\bibitem [{\citenamefont {Abbasi}\ \emph {et~al.}(2018)\citenamefont {Abbasi}
  \emph {et~al.}}]{TAmuons}%
  \BibitemOpen
  \bibfield  {author} {\bibinfo {author} {\bibfnamefont {R.~U.}\ \bibnamefont
  {Abbasi}} \emph {et~al.} (\bibinfo {collaboration} {Telescope Array
  Collaboration}),\ }\href {\doibase 10.1103/PhysRevD.98.022002} {\bibfield
  {journal} {\bibinfo  {journal} {Phys. Rev. D}\ }\textbf {\bibinfo {volume}
  {98}},\ \bibinfo {pages} {022002} (\bibinfo {year} {2018})}\BibitemShut
  {NoStop}%
\bibitem [{\citenamefont {Aab}\ \emph {et~al.}(2015)\citenamefont {Aab} \emph
  {et~al.}}]{MuonDeficit15}%
  \BibitemOpen
  \bibfield  {author} {\bibinfo {author} {\bibfnamefont {A.}~\bibnamefont
  {Aab}} \emph {et~al.} (\bibinfo {collaboration} {Pierre Auger
  Collaboration}),\ }\href {\doibase 10.1103/PhysRevD.91.032003} {\bibfield
  {journal} {\bibinfo  {journal} {Phys. Rev. D}\ }\textbf {\bibinfo {volume}
  {91}},\ \bibinfo {pages} {032003} (\bibinfo {year} {2015})}\BibitemShut
  {NoStop}%
\bibitem [{\citenamefont {{S. M\"uller for the Pierre Auger
  Collaboration}}(2019)}]{Muller18}%
  \BibitemOpen
  \bibfield  {author} {\bibinfo {author} {\bibnamefont {{S. M\"uller for the
  Pierre Auger Collaboration}}},\ }\href {\doibase
  10.1051/epjconf/201921002013} {\bibfield  {journal} {\bibinfo  {journal} {EPJ
  Web Conf.}\ }\textbf {\bibinfo {volume} {210}},\ \bibinfo {pages} {02013}
  (\bibinfo {year} {2019})}\BibitemShut {NoStop}%
\bibitem [{\citenamefont {Hayashida}\ \emph {et~al.}(1995)\citenamefont
  {Hayashida} \emph {et~al.}}]{AGASA}%
  \BibitemOpen
  \bibfield  {author} {\bibinfo {author} {\bibfnamefont {N.}~\bibnamefont
  {Hayashida}} \emph {et~al.},\ }\href@noop {} {\bibfield  {journal} {\bibinfo
  {journal} {J. Phys. G: Nucl. Part. Phys.}\ }\textbf {\bibinfo {volume}
  {21}},\ \bibinfo {pages} {1101} (\bibinfo {year} {1995})}\BibitemShut
  {NoStop}%
\bibitem [{\citenamefont {{D. Ivanov for the Pierre Auger Collaboration and the
  Telescope Array Collaboration}}(2017)}]{Ivanov17}%
  \BibitemOpen
  \bibfield  {author} {\bibinfo {author} {\bibnamefont {{D. Ivanov for the
  Pierre Auger Collaboration and the Telescope Array Collaboration}}},\ }in\
  \href@noop {} {\emph {\bibinfo {booktitle} {Proceedings of the 35th
  International Cosmic Ray Conference (ICRC)}}},\ Vol.\ \bibinfo {volume}
  {301}\ (\bibinfo {year} {2017})\ p.\ \bibinfo {pages} {498}\BibitemShut
  {NoStop}%
\bibitem [{\citenamefont {{Dembinski}}(2018)}]{Dembinski18}%
  \BibitemOpen
  \bibfield  {author} {\bibinfo {author} {\bibfnamefont {H.~P.}\ \bibnamefont
  {{Dembinski}}},\ }\href {\doibase 10.1016/j.astropartphys.2018.05.008}
  {\bibfield  {journal} {\bibinfo  {journal} {Astroparticle Physics}\ }\textbf
  {\bibinfo {volume} {102}},\ \bibinfo {pages} {89} (\bibinfo {year} {2018})},\
  \Eprint {http://arxiv.org/abs/1711.05737} {arXiv:1711.05737 [astro-ph.HE]}
  \BibitemShut {NoStop}%
\bibitem [{\citenamefont {Heck}\ \emph {et~al.}(1998)\citenamefont {Heck},
  \citenamefont {Knapp}, \citenamefont {Capdevielle}, \citenamefont {Schatz},\
  and\ \citenamefont {Thouw}}]{CORSIKA}%
  \BibitemOpen
  \bibfield  {author} {\bibinfo {author} {\bibfnamefont {D.}~\bibnamefont
  {Heck}}, \bibinfo {author} {\bibfnamefont {J.}~\bibnamefont {Knapp}},
  \bibinfo {author} {\bibfnamefont {J.~N.}\ \bibnamefont {Capdevielle}},
  \bibinfo {author} {\bibfnamefont {G.}~\bibnamefont {Schatz}}, \ and\ \bibinfo
  {author} {\bibfnamefont {T.}~\bibnamefont {Thouw}},\ }\href {\doibase
  10.5445/IR/270043064} {\emph {\bibinfo {title} {{CORSIKA: A Monte Carlo code
  to simulate extensive air showers}}}},\ \bibinfo {type} {Tech. Rep.}\
  \bibinfo {number} {FZKA-6019}\ (\bibinfo  {institution} {Forschungszentrum
  Karlsruhe},\ \bibinfo {year} {1998})\BibitemShut {NoStop}%
%%CITATION = FZKA-6019;%%
\bibitem [{\citenamefont {{C. Jui for the Telescope Array
  Collaboration}}(2015)}]{TAFlux}%
  \BibitemOpen
  \bibfield  {author} {\bibinfo {author} {\bibnamefont {{C. Jui for the
  Telescope Array Collaboration}}},\ }in\ \href@noop {} {\emph {\bibinfo
  {booktitle} {Proceedings of the 34th International Cosmic Ray Conference
  (ICRC)}}},\ Vol.~\bibinfo {volume} {34}\ (\bibinfo {year} {2015})\
  p.~\bibinfo {pages} {35}\BibitemShut {NoStop}%
\bibitem [{\citenamefont {{M. Takeda}}\ \emph {et~al.}(2003)\citenamefont {{M.
  Takeda}} \emph {et~al.}}]{Takeda2003}%
  \BibitemOpen
  \bibfield  {author} {\bibinfo {author} {\bibnamefont {{M. Takeda}}} \emph
  {et~al.},\ }\href@noop {} {\bibfield  {journal} {\bibinfo  {journal}
  {Astroparticle Physics}\ }\textbf {\bibinfo {volume} {19}},\ \bibinfo {pages}
  {447} (\bibinfo {year} {2003})}\BibitemShut {NoStop}%
\bibitem [{\citenamefont {{S. Yoshida}}\ \emph {et~al.}(1995)\citenamefont {{S.
  Yoshida}} \emph {et~al.}}]{Yoshida95}%
  \BibitemOpen
  \bibfield  {author} {\bibinfo {author} {\bibnamefont {{S. Yoshida}}} \emph
  {et~al.},\ }\href@noop {} {\bibfield  {journal} {\bibinfo  {journal}
  {Astroparticle Physics}\ }\textbf {\bibinfo {volume} {3}},\ \bibinfo {pages}
  {105} (\bibinfo {year} {1995})}\BibitemShut {NoStop}%
\bibitem [{\citenamefont {Brun}\ and\ \citenamefont {Rademakers}(1997)}]{ROOT}%
  \BibitemOpen
  \bibfield  {author} {\bibinfo {author} {\bibfnamefont {R.}~\bibnamefont
  {Brun}}\ and\ \bibinfo {author} {\bibfnamefont {F.}~\bibnamefont
  {Rademakers}},\ }\bibfield  {booktitle} {\emph {\bibinfo {booktitle} {{New
  computing techniques in physics research V. Proceedings, 5th International
  Workshop, AIHENP '96, Lausanne, Switzerland, September 2-6, 1996}}},\ }\href
  {\doibase 10.1016/S0168-9002(97)00048-X} {\bibfield  {journal} {\bibinfo
  {journal} {Nucl. Instrum. Meth.}\ }\textbf {\bibinfo {volume} {A389}},\
  \bibinfo {pages} {81} (\bibinfo {year} {1997})}\BibitemShut {NoStop}%
%%CITATION = NUIMA,A389,81;%%
\bibitem [{\citenamefont {Ivanov}(2012)}]{doctesisIvanov}%
  \BibitemOpen
  \bibfield  {author} {\bibinfo {author} {\bibfnamefont {D.}~\bibnamefont
  {Ivanov}},\ }\emph {\bibinfo {title} {{Energy spectrum measured by the
  telescope array surface detector}}},\ \href
  {https://rucore.libraries.rutgers.edu/rutgers-lib/38839/} {Ph.D. thesis},\
  \bibinfo  {school} {Rutgers U., Piscataway} (\bibinfo {year}
  {2012})\BibitemShut {NoStop}%
%%CITATION = INSPIRE-1513829;%%
\bibitem [{\citenamefont {{V. Verzi for the Pierre Auger
  Collaboration}}(2013)}]{EScaleAuger}%
  \BibitemOpen
  \bibfield  {author} {\bibinfo {author} {\bibnamefont {{V. Verzi for the
  Pierre Auger Collaboration}}},\ }in\ \href@noop {} {\emph {\bibinfo
  {booktitle} {{The Pierre Auger Observatory: Contributions to the 33rd
  International Cosmic Ray Conference (ICRC)}}}}\ (\bibinfo {year} {2013})\
  pp.\ \bibinfo {pages} {7--10}\BibitemShut {NoStop}%
\bibitem [{\citenamefont {{Abbasi}}\ \emph {et~al.}(2016)\citenamefont
  {{Abbasi}} \emph {et~al.}}]{EScaleTA}%
  \BibitemOpen
  \bibfield  {author} {\bibinfo {author} {\bibfnamefont {R.~U.}\ \bibnamefont
  {{Abbasi}}} \emph {et~al.},\ }\href {\doibase
  10.1016/j.astropartphys.2016.04.002} {\bibfield  {journal} {\bibinfo
  {journal} {Astroparticle Physics}\ }\textbf {\bibinfo {volume} {80}},\
  \bibinfo {pages} {131} (\bibinfo {year} {2016})}\BibitemShut {NoStop}%
\bibitem [{\citenamefont {{K. Shinozaki}}\ and\ \citenamefont {{M.
  Teshima}}(2004)}]{Shinozaki2004}%
  \BibitemOpen
  \bibfield  {author} {\bibinfo {author} {\bibnamefont {{K. Shinozaki}}}\ and\
  \bibinfo {author} {\bibnamefont {{M. Teshima}}},\ }\href@noop {} {\bibfield
  {journal} {\bibinfo  {journal} {Nuclear Physics B (Proc. Suppl.)}\ }\textbf
  {\bibinfo {volume} {136}},\ \bibinfo {pages} {18} (\bibinfo {year}
  {2004})}\BibitemShut {NoStop}%
\bibitem [{\citenamefont {{K. Shinozaki}}\ \emph {et~al.}(2001)\citenamefont
  {{K. Shinozaki}} \emph {et~al.}}]{Shinozaki2001}%
  \BibitemOpen
  \bibfield  {author} {\bibinfo {author} {\bibnamefont {{K. Shinozaki}}} \emph
  {et~al.},\ }in\ \href@noop {} {\emph {\bibinfo {booktitle} {Proceedings of
  the 27th International Cosmic Ray Conference (ICRC)}}},\ Vol.~\bibinfo
  {volume} {1}\ (\bibinfo {year} {2001})\ p.\ \bibinfo {pages}
  {346}\BibitemShut {NoStop}%
\bibitem [{\citenamefont {{K. Shinozaki}}\ \emph {et~al.}(2002)\citenamefont
  {{K. Shinozaki}} \emph {et~al.}}]{Shinozaki2002}%
  \BibitemOpen
  \bibfield  {author} {\bibinfo {author} {\bibnamefont {{K. Shinozaki}}} \emph
  {et~al.},\ }\href {\doibase 10.1086/341288} {\bibfield  {journal} {\bibinfo
  {journal} {The Astrophysical Journal Letters}\ }\textbf {\bibinfo {volume}
  {571}},\ \bibinfo {pages} {L117} (\bibinfo {year} {2002})}\BibitemShut
  {NoStop}%
\bibitem [{\citenamefont {{B{\"o}hlen}}\ \emph {et~al.}(2014)\citenamefont
  {{B{\"o}hlen}}, \citenamefont {{Cerutti}}, \citenamefont {{Chin}},
  \citenamefont {{Fass{\`o}}}, \citenamefont {{Ferrari}}, \citenamefont
  {{Ortega}}, \citenamefont {{Mairani}}, \citenamefont {{Sala}}, \citenamefont
  {{Smirnov}},\ and\ \citenamefont {{Vlachoudis}}}]{Fluka1}%
  \BibitemOpen
  \bibfield  {author} {\bibinfo {author} {\bibfnamefont {T.~T.}\ \bibnamefont
  {{B{\"o}hlen}}}, \bibinfo {author} {\bibfnamefont {F.}~\bibnamefont
  {{Cerutti}}}, \bibinfo {author} {\bibfnamefont {M.~P.~W.}\ \bibnamefont
  {{Chin}}}, \bibinfo {author} {\bibfnamefont {A.}~\bibnamefont {{Fass{\`o}}}},
  \bibinfo {author} {\bibfnamefont {A.}~\bibnamefont {{Ferrari}}}, \bibinfo
  {author} {\bibfnamefont {P.~G.}\ \bibnamefont {{Ortega}}}, \bibinfo {author}
  {\bibfnamefont {A.}~\bibnamefont {{Mairani}}}, \bibinfo {author}
  {\bibfnamefont {P.~R.}\ \bibnamefont {{Sala}}}, \bibinfo {author}
  {\bibfnamefont {G.}~\bibnamefont {{Smirnov}}}, \ and\ \bibinfo {author}
  {\bibfnamefont {V.}~\bibnamefont {{Vlachoudis}}},\ }\href {\doibase
  10.1016/j.nds.2014.07.049} {\bibfield  {journal} {\bibinfo  {journal}
  {Nuclear Data Sheets}\ }\textbf {\bibinfo {volume} {120}},\ \bibinfo {pages}
  {211} (\bibinfo {year} {2014})}\BibitemShut {NoStop}%
\bibitem [{\citenamefont {Ferrari}\ \emph {et~al.}(2005)\citenamefont
  {Ferrari}, \citenamefont {Sala}, \citenamefont {Fassò},\ and\ \citenamefont
  {Ranft}}]{Fluka2}%
  \BibitemOpen
  \bibfield  {author} {\bibinfo {author} {\bibfnamefont {A.}~\bibnamefont
  {Ferrari}}, \bibinfo {author} {\bibfnamefont {P.~R.}\ \bibnamefont {Sala}},
  \bibinfo {author} {\bibfnamefont {A.}~\bibnamefont {Fassò}}, \ and\ \bibinfo
  {author} {\bibfnamefont {J.}~\bibnamefont {Ranft}},\ }\href {\doibase
  10.5170/CERN-2005-010} {\emph {\bibinfo {title} {{FLUKA: A multi-particle
  transport code (program version 2005)}}}},\ CERN Yellow Reports: Monographs\
  (\bibinfo  {publisher} {CERN},\ \bibinfo {address} {Geneva},\ \bibinfo {year}
  {2005})\BibitemShut {NoStop}%
\bibitem [{\citenamefont {{F. Fenu for the Pierre Auger
  Collaboration}}(2017)}]{Fenu17}%
  \BibitemOpen
  \bibfield  {author} {\bibinfo {author} {\bibnamefont {{F. Fenu for the Pierre
  Auger Collaboration}}} (\bibinfo {collaboration} {Pierre Auger}),\ }in\ \href
  {\doibase 10.22323/1.301.0486} {\emph {\bibinfo {booktitle} {{The Pierre
  Auger Observatory: Contributions to the 35th International Cosmic Ray
  Conference (ICRC)}}}}\ (\bibinfo {year} {2017})\ pp.\ \bibinfo {pages}
  {9--16}\BibitemShut {NoStop}%
%%CITATION = INSPIRE-1618413;%%
\bibitem [{\citenamefont {{National Centers for Environmental Information,
  National Oceanic and Atmospheric Administration}}()}]{Geomag}%
  \BibitemOpen
  \bibfield  {author} {\bibinfo {author} {\bibnamefont {{National Centers for
  Environmental Information, National Oceanic and Atmospheric
  Administration}}},\ }\href {http://www.ngdc.noaa.gov/geomag/} {}\bibinfo
  {note} {{accessed September 2, 2019}}\BibitemShut {NoStop}%
\bibitem [{\citenamefont {{M. Kobal for the Pierre Auger
  Collaboration}}(2001)}]{Kobal2001}%
  \BibitemOpen
  \bibfield  {author} {\bibinfo {author} {\bibnamefont {{M. Kobal for the
  Pierre Auger Collaboration}}},\ }\href {\doibase
  10.1016/S0927-6505(00)00158-4} {\bibfield  {journal} {\bibinfo  {journal}
  {Astroparticle Physics}\ }\textbf {\bibinfo {volume} {15}},\ \bibinfo {pages}
  {259} (\bibinfo {year} {2001})}\BibitemShut {NoStop}%
\end{thebibliography}%

\end{document}